\documentclass[showpacs, pre, superscriptaddress,nofootinbib, twocolumn]{revtex4-1}
\usepackage{amsmath,amssymb,amsthm,mathrsfs,amsopn}
\usepackage{graphicx}
\usepackage{subfigure}
\usepackage{booktabs}
\usepackage{array}
\usepackage{tcolorbox} 
\tcbset{colback=blue!5!white,colframe=blue!50!black, left=0mm,right=0mm,top=0mm,bottom=0mm, box align=center, halign=center} 
\usepackage{tikz} 
\usepackage{enumitem}
\usepackage{fancyhdr} 

\begin{document}

\title{Reconstruction scheme for excitatory and inhibitory dynamics with quenched disorder: application to zebrafish imaging}
\author{Lorenzo Chicchi}
\thanks{Corresponding author}
\email{chicchi.lorenzo@gmail.com}
\affiliation{Department of Physics and Astronomy, University of Florence, Sesto Fiorentino, Florence, Italy}
\affiliation{CSDC, University of Florence, Sesto Fiorentino, Florence, Italy}

\author{Gloria Cecchini}
\affiliation{Department of Physics and Astronomy, University of Florence, Sesto Fiorentino, Florence, Italy}
\affiliation{CSDC, University of Florence, Sesto Fiorentino, Florence, Italy}

\author{Ihusan Adam}
\affiliation{Department of Physics and Astronomy, University of Florence, Sesto Fiorentino, Florence, Italy}
\affiliation{CSDC, University of Florence, Sesto Fiorentino, Florence, Italy}
\affiliation{Department of Information Engineering, University of Florence, Florence, Italy}

\author{Giuseppe de Vito}
\affiliation{European Laboratory for Non-Linear Spectroscopy, Sesto Fiorentino, Florence, Italy}
\affiliation{Department of Neuroscience, Psychology, Drug Research and Child Health, University of Florence, Florence, Italy}

\author{Roberto Livi}
\affiliation{Department of Physics and Astronomy, University of Florence, Sesto Fiorentino, Florence, Italy}
\affiliation{CSDC, University of Florence, Sesto Fiorentino, Florence, Italy}
\affiliation{INFN Sezione di Firenze, Sesto Fiorentino, Florence, Italy}

\author{Francesco Saverio Pavone}
\affiliation{Department of Physics and Astronomy, University of Florence, Sesto Fiorentino, Florence, Italy}
\affiliation{European Laboratory for Non-Linear Spectroscopy, Sesto Fiorentino, Florence, Italy}
\affiliation{National Institute of Optics, National Research Councily, Sesto Fiorentino, Florence, Italy}

\author{Ludovico Silvestri}
\affiliation{Department of Physics and Astronomy, University of Florence, Sesto Fiorentino, Florence, Italy}
\affiliation{European Laboratory for Non-Linear Spectroscopy, Sesto Fiorentino, Florence, Italy}
\affiliation{National Institute of Optics, National Research Councily, Sesto Fiorentino, Florence, Italy}

\author{Lapo Turrini}
\affiliation{Department of Physics and Astronomy, University of Florence, Sesto Fiorentino, Florence, Italy}
\affiliation{European Laboratory for Non-Linear Spectroscopy, Sesto Fiorentino, Florence, Italy}

\author{Francesco Vanzi}
\affiliation{European Laboratory for Non-Linear Spectroscopy, Sesto Fiorentino, Florence, Italy}
\affiliation{Department of Biology, University of Florence, Sesto Fiorentino, Florence, Italy}

\author{Duccio Fanelli}
\affiliation{Department of Physics and Astronomy, University of Florence, Sesto Fiorentino, Florence, Italy}
\affiliation{CSDC, University of Florence, Sesto Fiorentino, Florence, Italy}
\affiliation{INFN Sezione di Firenze, Sesto Fiorentino, Florence, Italy}

\date{\today}

\begin{abstract}
An inverse procedure is developed and tested to recover functional and structural information from global signals of brains activity. The method assumes a leaky-integrate and fire model with excitatory and inhibitory neurons, coupled via a directed network. Neurons are endowed with a heterogenous current value, which sets their associated dynamical regime. By making use of a heterogenous mean-field approximation, the method seeks to reconstructing from global activity patterns the distribution of  in-coming degrees, for both excitatory and inhibitory neurons, as well as the distribution of the assigned currents. The proposed inverse scheme is first validated against synthetic data. Then, time-lapse acquisitions of a zebrafish larva recorded with a two-photon light sheet microscope are used as an input to the reconstruction algorithm. A power law distribution of the in-coming connectivity of the excitatory neurons is found. Local degree distributions are also computed by segmenting the whole brain in sub-regions traced from annotated atlas.
\end{abstract}

\maketitle

\section{Introduction}

The unmatched ability of the brain to cope with an extraordinarily large plethora of complex tasks, carried out in parallel, ultimately resides in the intricate web of interlinked connections which define the architecture of the embedding neurons’ network. Structural and functional information, as inferred from direct measurements of neuronal activity, under different experimental conditions, are fundamental pieces of a jigsaw puzzle of how the brains works, from simple organisms to more complicated creatures, across phylogenetic scales. Suitable methods have been developed which build on  statistical mechanics tools, e.g. maximum entropy principles \cite{bialek,cocco}, to resolve the functional map that orchestrates the coordinated firing of neurons dislocated in different portions of the brain. However, other sources of heterogeneity in the brain should be accounted for as well. Neuronal excitability, namely the ability of neurons to respond to external inputs, is finely controlled through inhibitory/excitatory balance \cite{deghani, neuromod}. Furthermore, individual neurons can display a variable degree of inherent excitability, a source of spatial quenched disorder which reflects back in the ensuing activation patterns. 

Motivated by this, in \cite{adam_HMF} we proposed and tested against both synthetic and real data, an inverse scheme to quantify the statistics of neurons'  excitability, while inferring, from global activity measurements, the, a priori unknown, distribution of network connectivities. The method employs an extended model of Leaky-Integrate and Fire (LIF) neurons,  with short-term plasticity. Only excitatory neurons are accounted for in  \cite{adam_HMF}. These are assumed to be coupled via a directed network and display a degree of heterogeneity in the associated current, which sets the  firing regime in which a neuron operates. The inverse scheme builds on the celebrated Heterogenous  Mean-Field (HMF) approximation \cite{barrat2008,dorogovtsev2008,pastor2001,vespignani2012} and seeks to recover the distribution of 
the (in-degree) connectivity of (excitatory) neurons, concurrently with the distribution of the assigned currents, denoted by $a$. The HMF approximation was previously employed in \cite{diVolo1,diVolo2,diVolo3, AFCI}  to reconstruct the topology of an underlying network from artificially generated data, meant to mimic neuronal signals. In \cite{adam_HMF} the approach was generalized so as to account for the dynamical heterogeneity, as stemming from the intrinsic degree of individual neurons’ excitability. As mentioned above, individual excitability acts as a key component of the dynamics and yields irregular patterns of activity like those displayed in real measurements. The reconstruction scheme was applied  in \cite{adam_HMF} to longitudinal wide-field fluorescence microscopy data of cortical functionality in groups of awake mice and enabled us to identify altered distributions in neuron excitability immediately after the stroke, and in agreement with earlier observation \cite{HyperExcitability1,HyperExcitability2,HyperExcitability3}. Conversely, rehabilitation allowed to recover a distribution similar to pre-stroke conditions.

The goal of this work is to push forward the reconstruction algorithm by accounting for the simultaneous presence of both excitatory and inhibitory neurons, in a refined variant of the inversion scheme proposed in \cite{adam_HMF}. Notice that already in \cite{diVolo3} intertangled families of excitatory and inhibitory neurons have been considered, but only in a simplified setting where currents were assumed to be homogeneous. Relaxing this ansatz proves however mandatory when aiming at bridging the  gap between theory and experiments, a challenge that we shall hereafter tackle. In particular, we will recast the dynamics of the examined LIF model in a reduced setting by grouping in different classes (excitatory and inhibitory) neurons which bear distinct values of the current $a$ and of the connectivity $k$. 

Our extended inverse method aims at computing the distribution of the currents, as well as the distributions of the connectivities, for respectively excitatory and inhibitory neurons, via an iterative scheme which self-consistently identifies the classes of neurons needed to interpolate the global activity field, supplied as an input to the algorithm. First, the performance of the method is evaluated in silico, against synthetically generated data. We then move forward to considering a direct application of the developed technique to custom-made two-photon Light Sheet (2PLS) microscope, optimized for high-speed (1 Hz) volumetric imaging of zebrafish larva (ZL, {\it Danio rerio}). Near infrared (NIR) light is used for excitation, covering a wavelength range that is not visible to the larva in order not to induce unwanted visual responses. Hence 2PLS microscopy allows to record whole-brain activity with high temporal and spatial resolution, by preventing undesired external bias \cite{wolf2015whole,de2020twoa,de2020twob}. The experimental input is processed with a properly devised methodology to return a spatially resolved raster plot for the spiking activity of neurons over time, which provides the ideal input for the reconstruction method to work. A power law distribution of the in-coming connectivity of excitatory neurons is found, which is robust over a significant range of the imposed fraction of inhibitory neurons. Local degree distributions are also recovered by partitioning the whole brain in bound sub-domains, traced from annotated atlas \cite{randlett2015whole,kunst2019cellular}. When manipulating experimental data one cannot distinguish among the contributions resulting from different neurons (excitatory vs. inhibitory). A procedure is however developed which allows for the degree distribution of the excitatory neurons to be determined, while accounting for the role exerted by the population of inhibitory ones.

The paper is organized as follows. Next section is devoted to introducing the reconstruction scheme and to challenge its performance against synthetically produced data. We will then turn to presenting the experimental platform and discuss the details that relate to data processing. Processed data are supplied as an input to the mathematical reconstruction scheme to yield the results which are presented in Section \ref{Sec_data}. Finally we will sum up and draw our conclusions.

\section{General mathematical framework}

\subsection{The model}
\label{SecModel}
We use a Leaky Integrate and Fire (LIF) model to mimic the dynamics of individual neurons. Consider a pool of $N$ LIF neurons and denote by $v_i(t)$ the membrane potential of neuron $i$. Further, label with $I^{syn}_i(t)$ the synaptic current due to the incoming connections with other neurons of the collection. 

The dynamics of the membrane potential is hence ruled by the following equation
\begin{equation}
\frac{dv_i(t)}{dt}=a_i-v_i(t)+I^{syn}_i(t)
\label{MemPot}
\end{equation}	
where $a_i$ stands for the external input current of  neuron $i$. This is a crucial quantity, as it sets the dynamics of the corresponding neuron, in the uncoupled
regime. The critical value $a_i = 1$  separates between quiescent and active (spiking) regimes. When $v_i$ reaches the threshold value $v_{th}$, neuron $i$ emits a spike and the membrane potential $v_i$ is reset to the base value $v_r$.
Following \cite{diVolo3}, the membrane potential is rescaled by a suitable amount to have the spike threshold set at $v_{th}=1$ and the rest potential at $v_r=0$.

The Tsodyks, Uziel, and Markram model \cite{TUM1,TUM2} is assumed to describe the interactions among neurons, i.e. their coupling dynamics. More specifically, the dynamics of a synapse is expressed in terms of fractions of three different neurotransmitter states: active ($y_{ij}(t)$), available ($x_{ij}(t)$) and inactive ($z_{ij}(t)$), where $i$ and $j$ stand respectively for post-synaptic and pre-synaptic neurons. The obvious constraint $x_{ij}(t)+y_{ij}(t)+z_{ij}(t)=1$ applies, for any time $t$.

During a spike of the pre-synaptic neuron, a fraction $u_{ij}$ of neurotransmitters in the available state is activated.
The time evolution of the three states
\begin{subequations}\label{Nuerotrans}
\begin{align} 
\dot{y}_{ij}(t)&=-\frac{y_{ij}(t)}{\tau_{in}}+u_{ij}x_{ij}S_j(t) \label{Nuerotrans1}\\
\dot{z}_{ij}(t)&=-\frac{z_{ij}(t)}{\tau_{r}^{i}}+\frac{y_{ij}(t)}{\tau_{in}} \label{Nuerotrans2}\\
x_{ij}(t)&+y_{ij}(t)+z_{ij}(t)=1 \label{Nuerotrans3}
\end{align}
\end{subequations}
takes as input the spike train $S_j(t)=\sum_m \delta(t-t_j^*(m))$ of pre-synaptic neuron $j$ emitting its \textit{n}-th spike at time $t_j(n)$ \cite{TUM}.
The time is rescaled to the membrane time constant $\tau_m=30ms$. The time constants are set to $\tau_{in}=0.2$, $\tau_r^i=3.4$ if the post-synaptic neuron $i$ is inhibitory, or $\tau_r^i=26.6$ if it is excitatory \cite{diVolo3}.

If the post-synaptic neuron $i$ is excitatory, the fraction $u_{ij}(t)$ of neurotransmitters is set to the constant value of $U=0.5$. Otherwise, $u_{ij}(t)$ evolves in time
\begin{equation}
\dot{u}_{ij}(t)=-\frac{u_{ij}}{\tau_f}+U_f(1-u_{ij}(t))S_j(t)
\label{NuerotransFraction}
\end{equation}
where $\tau_f=33.25$ and $U_f=0.08$ \cite{TUM,diVolo3}.

Equations (\ref{Nuerotrans}) are coupled to Eq.~(\ref{MemPot}) via the synaptic current
\begin{equation}
I^{syn}_i(t)=\frac{g}{N}\sum_{j\ne i}A_{ij}y_{ij}(t)
\end{equation}
where $g$ is the coupling parameter and $A_{ij}$ stands for the elements of the network adjacency matrix $A$. The matrix entry is $A_{ij}=1$, if a link exists which goes from $j$ to $i$, and provided $j$ is an excitatory neuron. Conversely, $A_{ij}=-1$ if the starting node $j$ identifies inhibitory neuron. On the other hand, if $A_{ij}=0$ nodes $i$ and $j$ are not directly connected. 

Equations~(\ref{Nuerotrans}) can be cast in a more compact form so as to favour insight into the inspected processes and reduce the associated computational costs. To achieve this, we first notice that Eqs.~(\ref{Nuerotrans}) only depend on the pre-synaptic neuron $j$ and the characteristics of the post-synaptic neuron $i$. Equation~(\ref{Nuerotrans1}) can hence be split into two distinct equations:  

\begin{equation}
\begin{split}
\dot{y}_j^E(t)&=-\frac{y_j^E(t)}{\tau_{in}}+u_j^E x_j^E S_j(t)\\ \dot{y}_j^I(t)&=-\frac{y_j^I(t)}{\tau_{in}}+u_j^I x_j^I S_j(t),
\end{split}
\end{equation}
where the apexes $E$ (Excitatory)  and $I$ (Inhibitory) reflect the specificity of the target neuron. Similar arguments apply to Eqs. (\ref{Nuerotrans2}), (\ref{Nuerotrans3}), and (\ref{NuerotransFraction}).

For each type of synapse, the average field, i.e. the fraction of neurotransmitters in the active state, is calculated as
\begin{equation}
\begin{split}
Y_{EI}(t)&=\frac{1}{f_IN}\sum_{i\in \mathcal{I}}y_i^E(t)\\
Y_{EE}(t)&=\frac{1}{(1-f_I)N}\sum_{i\in \mathcal{E}}y_i^E(t)\\
Y_{II}(t)&=\frac{1}{f_IN}\sum_{i\in \mathcal{I}}y_i^I(t)\\
Y_{IE}(t)&=\frac{1}{(1-f_I)N}\sum_{i\in \mathcal{E}}y_i^I(t)
\end{split}
\label{fieldsYEE}
\end{equation}
where $\mathcal{E}$ and $\mathcal{I}$ are the ensemble of excitatory and inhibitory neurons, respectively; $f_I$ stands for the fraction of inhibitory neurons in the network. The global fields are defined as
\begin{equation}
\begin{split}
Y_E(t)&=-f_IY_{EI}(t)+(1-f_I)Y_{EE}(t)\\
Y_I(t)&=-f_IY_{II}(t)+(1-f_I)Y_{IE}(t).
\end{split}
\label{fieldsYE}
\end{equation}

\subsection{The heterogeneous mean field ansatz.}
The model described in the previous section is reformulated here in terms of a Heterogeneous Mean Field (HMF) approximation.
The original neurons are classified according to their characteristics. 
More specifically, neurons of the same type (excitatory or inhibitory) and with the same incoming connectivity $k$ and external current $a$ are considered identical.
Therefore, $L\times M$ equivalence classes are defined, where $L$ ($M$) is the number of sub-intervals in which the value range of $k$ ($a$) has been divided.
Moreover, we assume that neurons in the class $k$ are subjected to a synaptic current proportional to their in-degree, i.e.,
\begin{equation}
\begin{split}
&\frac{g}{N}\sum_{j}A_{ij}y_j^E(t)\longrightarrow g\tilde{k}Y_E(t)\\
&\frac{g}{N}\sum_{j}A_{ij}y_j^I(t)\longrightarrow g\tilde{k}Y_I(t)
\end{split}
\end{equation}
for the excitatory and inhibitory neurons, respectively.
Following this assumption, the model can be rewritten as
\begin{equation}
\begin{split}
\dot{v}_{k,a}^{E}(t)&=a-v_{k,a}^E(t)+g\tilde{k}Y_E(t)\\
\dot{v}_{k,a}^{I}(t)&=a-v_{k,a}^I(t)+g\tilde{k}Y_I(t)\\
\\
\dot{y}_{k,a}^{(\dagger,*)}(t)&=-\frac{y_{k,a}^{(\dagger,*)}(t)}{\tau_{in}}+u_{k,a}^{(\dagger,*)}(t)x_{k,a}^{(\dagger,*)}(t)S_{k,a}^{*}(t)\\
\dot{z}_{k,a}^{(\dagger,*)}(t)&=\frac{y_{k,a}^{(\dagger,*)}(t)}{\tau_{in}}-\frac{z_{k,a}^{(\dagger,*)}(t)}{\tau_{r}^{\dagger}}\\
x_{k,a}^{(\dagger,*)}(t)&+y_{k,a}^{(\dagger,*)}(t)+z_{k,a}^{(\dagger,*)}(t)=1
\end{split}
\label{HMF}
\end{equation}
where $(\dagger,*)$ identify all possible pairs of  post-synaptic and pre-synaptic neurons. 
Denote by $P_E(k)$ and $P_I(k)$ the in-degree distributions for the excitatory and inhibitory neurons, respectively, and $P(a)$ the external current distribution.
Equations (\ref{HMF}) can be closed by the consistency relations
\begin{equation}
\tilde{Y}_{\dagger *}(t)=\int_{k,a}P_{*}(k)P(a)y_{k,a}^{\dagger *}(t).
\label{daggerY}
\end{equation}
Taking in account the discretization of the defined classes of equivalence, Eq. (\ref{daggerY}) turns into
\begin{equation}
\tilde{Y}_{\dagger *}(t)=\sum_{l=1}^{L}\sum_{m=1}^{M}P_{*}(k_l)P(a_m) y_{k_l,a_m}^{\dagger *}(t),
\label{Close}
\end{equation}
where we have implicitly introduced the discrete counterpart of the continuous probability distributions $\mathbf{P_{*}(k)}=(P_{*}(k_1),P_{*}(k_2),...,P_{*}(k_L)) $ and $\mathbf{P(a)}=(P(a_1),P(a_2),..,P(a_M))$.

\subsection{Reconstruction scheme}\label{ReconstrucionScheme}

\begin{figure*}[!ht]
\centering
\includegraphics[width=0.9\textwidth]{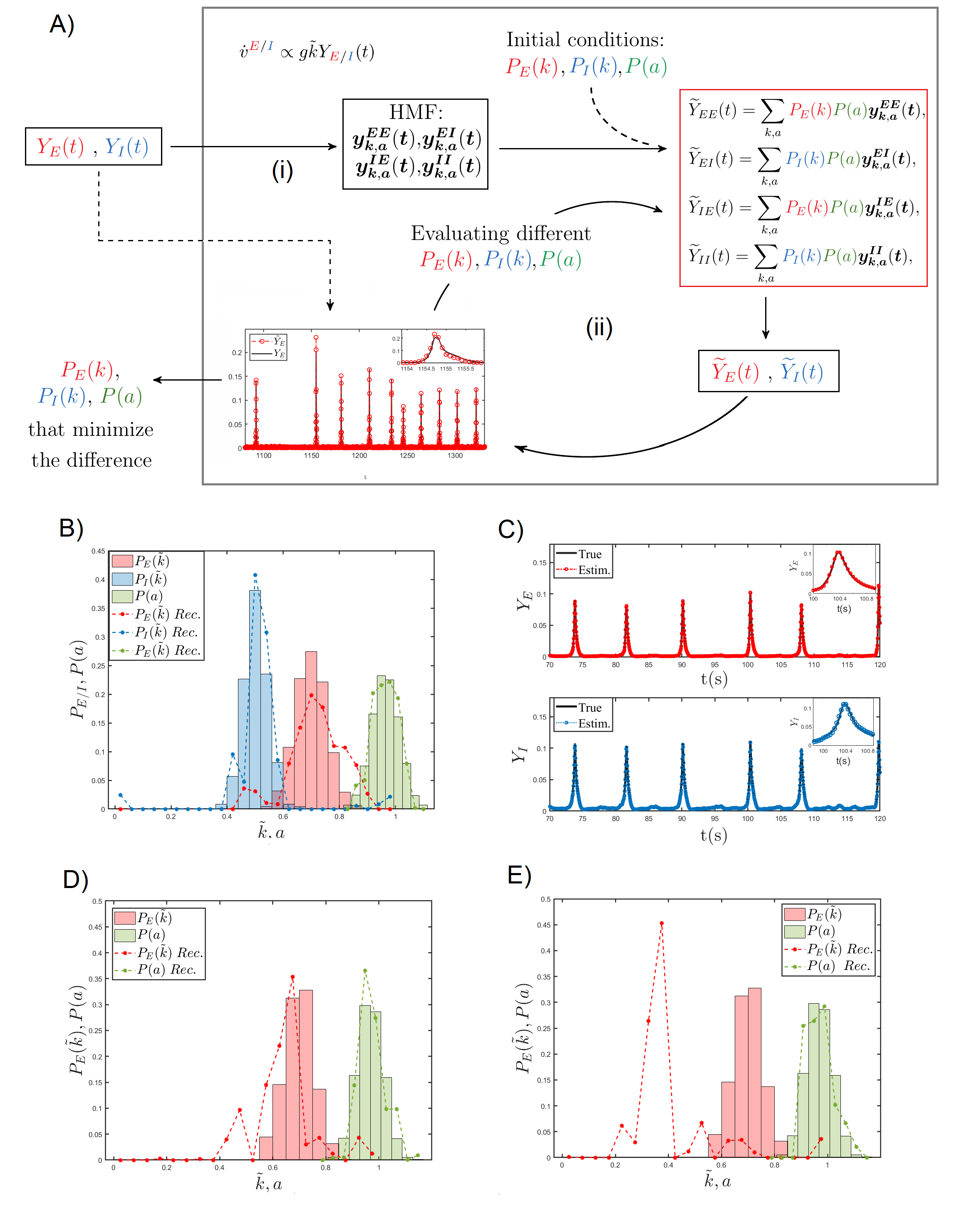}
\caption{\footnotesize A) Schematic outline of the reconstruction procedure. The global fields $Y_E(t)$ and $Y_I(t)$ constitute the inputs of the model in the HMF approximation (i). Different choices for the probability distributions $P_E(k)$, $P_I(k)$, and $P(a)$ are iteratively tested in order to find the best match between the input fields and the reconstructed fields, as obtained by using the equations displayed in the red box (ii). B) Outcome of the reconstruction procedure: the true probability distributions of a synthetic network are compared with those obtained with the proposed reconstruction method. A random network with $N=5000$ is considered here. The fraction of inhibitory neurons is set to $f_I=0.05$. The number of classes defined in the HMF approximation for the in-degree and the external current is $L=250$ and $M=250$ respectively. C) Comparison between the true global fields and the ones obtained via the reconstructed distributions. The plot in the inset is a zoom in of a peak. D-E) Outputs of the reconstruction are compared with the true external current probability distribution $P(a)$ and the true in-degree distribution $P_E(k)$ for the excitatory neurons of the same network; the network is made of $N=1000$ neurons of which a fraction $f_I=0.2$ are inhibitors. In the HMF approximation one hundred classes have been defined for both the in-degree and external current, namely,  $L=100$ and $M=100$. In D) the correct fraction of inhibitory neurons is taken into account, while in E) the inhibitory neuron effects are not considered.}
\label{Pannello1}
\end{figure*}

In this section, we set up a general reconstruction scheme for recovering the a priori unknown distributions for the in-degree $P_E(k)$ and $P_I(k)$, as well as the external current $P(a)$. This is achieved by interpolating the available global fields  $Y_E(t)$ and $Y_I(t)$, under the simplified HMF descriptive framework. The main steps of the reconstruction algorithm are schematically depicted in Fig. \ref{Pannello1}A.

As already mentioned, we assume the global fields $Y_E(t)$ and $Y_I(t)$ to be given. Then we integrate Eqs.~(\ref{HMF}), by using these global fields $Y_E(t)$ and $Y_I(t)$ as inputs to the model. 
The equations are initialized with variables ($v_{k,a}(t_0), y_{k,a}^{(\dagger,*)}(t_0), z_{k,a}^{(\dagger,*)}(t_0)$) randomly drawn from a uniform distribution, and generated so as to respect the constraints $v^{\dagger}_{k,a}(t_0)<1$ and $y_{k,a}^{(\dagger,*)}(t_0) + z_{k,a}^{(\dagger,*)}(t_0)<1$.
Forced by the external fields $Y_E(t)$ and $Y_I(t)$, the governing equations are integrated forward and the variables $y_{k,a}^{\dagger *}(t)$ are stored for each class $(k,a)$, type of synapse $(\dagger,*)$, and time $t$.
This process is repeated for $H$ independent realizations of the initial conditions.
The average fraction of neurotransmitters in the active state, for each class $(k,a)$ and synapse type, is computed at any time of observation $t$, i.e., $\left\langle y_{k,a}^{(\dagger,*)}(t)\right\rangle=1/H\sum_{h=1}^H\big(y_{k,a}^{(\dagger,*)}(t)\big)_h$.

Then, the approximated global fields $\tilde{Y}_{\dagger *}$ are calculated, via Eq. (\ref{Close}), for an initial guess of the distributions $\mathbf{P_{E}(k)}$, $\mathbf{P_{I}(k)}$, and $\mathbf{P(a)}$. These latter are then recursively modified so as to improve the correspondence between the approximated fields and their true homologues $Y_{\dagger *}$. 

Formally, for each $(\dagger, *)\in\{EE,EI,IE,II\}$, we aim at minimizing the function
\begin{equation}
F^{\dagger *}(\mathbf{P_{*}(k),P(a)})=\sum_t \lvert Y_{\dagger *}(t)-\sum_{k,a}\mathbf{P_{*}(k)P(a)}\left\langle y_{k,a}^{\dagger *}(t)\right\rangle\rvert ^2.
\label{FunctionMin}
\end{equation}
Note that the arguments of the above function are the target probability distributions $\mathbf{P_{*}(k)} $ and $\mathbf{P(a)}$ that one aims at inferring.

The iterative algorithm operates as follows. The distribution $\mathbf{P(a)}$ is initially frozen to a given profile and the quantities $y_{k_l}^{\dagger *}(t)=\sum_{m=1}^{M} P(a_m) \left\langle y_{k_l,a_m}^{\dagger *}(t)\right\rangle$ are hence evaluated. The inverse problem yields therefore 
\begin{equation}
\begin{pmatrix}
1/\Delta k\\
Y_{\dagger *}(t_1)\\
Y_{\dagger *}(t_2)\\
Y_{\dagger *}(t_3)\\
\vdots
\end{pmatrix}\approx
\begin{pmatrix}
1 & 1 & 1 & \dots\\
y_{k_1}^{\dagger *}(t_1) & y_{k_2}^{\dagger *}(t_1) & y_{k_3}^{\dagger *}(t_1) & \dots \\
y_{k_1}^{\dagger *}(t_2) & y_{k_2}^{\dagger *}(t_2) & y_{k_3}^{\dagger *}(t_2) &\dots \\
y_{k_1}^{\dagger *}(t_3) & y_{k_2}^{\dagger *}(t_3) & y_{k_3}^{\dagger *}(t_3) & \dots \\
\vdots & \vdots & \vdots & \ddots
\end{pmatrix}
\begin{pmatrix}
P_{*}(k_1)\\
P_{*}(k_2)\\
P_{*}(k_3)\\
\vdots
\end{pmatrix},
\end{equation}
where the first row reflects the normalization condition. 
The problem is hence reduced to a linear system that can be readily solved to obtain a first estimate of the distributions $\mathbf{P_{E}(k)}$ and $\mathbf{P_{I}(k)}$. 

As second step, the in-degree distributions $\mathbf{P_{E}(k)}$ and $\mathbf{P_{I}(k)}$ are fixed to the solutions found at the previous iteration.
Similar to step one, we evaluate the quantities  $y_{a_m}^{\dagger *}(t)=\sum_{l=1}^{L} P_{*}(k_l) \left\langle y_{k_l,a_m}^{\dagger *}(t)\right\rangle$ and formulate the linear problem
\begin{equation}
\begin{pmatrix}
1/\Delta a\\
Y_{\dagger *}(t_1)\\
Y_{\dagger *}(t_2)\\
Y_{\dagger *}(t_3)\\
\vdots
\end{pmatrix}\approx
\begin{pmatrix}
1 & 1 & 1 & \dots\\
y_{a_1}^{\dagger *}(t_1) & y_{a_2}^{\dagger *}(t_1) & y_{a_3}^{\dagger *}(t_1) & \dots \\
y_{a_1}^{\dagger *}(t_2) & y_{a_2}^{\dagger *}(t_2) & y_{a_3}^{\dagger *}(t_2) &\dots \\
y_{a_1}^{\dagger *}(t_3) & y_{a_2}^{\dagger *}(t_3) & y_{a_3}^{\dagger *}(t_3) & \dots \\
\vdots & \vdots & \vdots & \ddots
\end{pmatrix}
\begin{pmatrix}
P(a_1)\\
P(a_2)\\
P(a_3)\\
\vdots
\end{pmatrix}
\end{equation}
The above problem can be solved to obtained an updated estimate for the $\mathbf{P(a)}$. 
The overall procedure, consisting of two nested steps, is iterated until a maximum number of allowed cycles is reached, or, alternatively, the stopping criterion is eventually met.

Before proceeding in the analysis, we introduce a slightly modified notation. The in-degree $k$ is normalized to the size of the network $N$. In formulae we will set $\tilde{k}=k/N$, with $\tilde{k}\in[0,1]$.
From hereon the distributions that constitute the target of the reconstruction scheme will be hence expressed as function of $\tilde{k}$, instead of $k$.

\section{Application to data}

\subsection{Synthetic data}
In this section we test the proposed reconstruction protocol against synthetic data. The reconstruction scheme was successfully validated on synthetic data for the case of homogeneous external current in \cite{diVolo3,adam_HMF}.
Here, we test the reconstruction method on synthetic networks of excitatory and inhibitory neurons, assuming a quenched distribution of heterogeneous external currents. To this end we generate a random graph with $N$ nodes whose structural characteristic is  contained in the signed $N \times N$ adjacency matrix $A$,
which specifies the existence of (directed) links among pairs of adjacent nodes. Following the convention introduced above, negative entries ($A_{ij}=-1$) indicate that the starting node ($j$) is of the inhibitory type, whereas for positive elements ($A_{ij}=1$) $j$ belongs to the family of excitatory neurons. The network generation procedure is conceived so as to return a bell shaped distribution for both $P_E(\tilde{k})$ and $P_I(\tilde{k})$ (see Fig.~ \ref{Pannello1}B). Quenched disorder in the input
currents is introduced, the assigned currents being distributed according to a uni-modal profile $P(a)$ (see Fig.~ \ref{Pannello1}B). These are the exact distributions that we eventually seek to recover via the aforementioned reconstruction algorithm. Note that the domain of definition of $P(a)$ includes the bifurcation value $a = 1$.

With this setting, Eqs. (\ref{MemPot}) and (\ref{Nuerotrans}) are integrated and the fields $Y_E$ and $Y_I$ are calculated by using Eqs. (\ref{fieldsYEE}) and (\ref{fieldsYE}).
This is possible because we have access to all information which pertain to the network architecture and to the 
heterogenous collection of randomly generated currents.

The recorded global fields $Y_E$ and $Y_I$ are used as inputs to the reconstruction scheme presented in the previous section. 
Figure \ref{Pannello1}B shows the comparison between true and estimated distributions, at the end of the reconstruction procedure and for one generation of the synthetic network. By inserting the estimated distributions into Eq. (\ref{Close}), we obtain the global fields $\tilde{Y}_E$ and $\tilde{Y}_I$. 
The comparison between estimated ($\tilde{Y}_E,\tilde{Y}_I$) and true ($Y_E,Y_I$) global fields, as obtained by working on the index space, is presented in Fig. \ref{Pannello1}C. The agreement is excellent for both the inhibitory and the excitatory components.

When working with experimental data, however, one cannot  isolate the contributions stemming from different neurons, grouped according to their specific traits (excitatory vs. inhibitory). This implies that the sums in Eq. (\ref{fieldsYEE}), i.e. the input to the envisaged reconstruction scheme, cannot be in general accessed, as it was instead the case when working in the framework of the synthetic model considered above. To overcome this intrinsic limitation, we  propose and test an alternative route, which performs adequately well when challenged against synthetic data. The idea is to propose an approximated version of the input Eqs. (\ref{fieldsYEE}). To this end, we extend the sums which run on the excitatory neurons to all neurons and compute, under this  approximation, the fields $Y_{EE}$ and $Y_{IE}$. To validate this hypothesis we operate with synthetic networks with unclassified neurons. It can be shown that the approximation for the fields $Y_{EE},Y_{IE}$ correctly describes $Y_{EE}$, but not $Y_{IE}$. This conclusion is supported by systematic numerical investigations (data not shown), that we carried out by varying the relative proportion of excitatory and inhibitory neurons. Building on the above, we therefore write:

\begin{equation}
	Y_{EE}(t)\approx \frac{1}{N}\sum_{i=1}^{N}y_i^{E}(t),
	\label{ApproxYee}
\end{equation}
which enables us to compute the approximated field $Y_{EI}$ as
\begin{equation}
\begin{split}
Y_{EI}(t)&=\frac{1}{f_IN}\sum_{i\in\mathcal{I}}y_i^E(t)=\\
&=\frac{1}{f_IN}\bigg(\sum_{i=1}^Ny_i^E(t)-\sum_{i\in\mathcal{E}}y_i^E(t)\bigg)=\\
&=\frac{1}{f_IN}\sum_{i=1}^Ny_i^E(t)-\frac{1}{f_IN}\frac{(1-f_I)}{(1-f_I)}\sum_{i\in\mathcal{E}}y_i^E(t)=\\
&=\frac{1}{f_IN}\sum_{i=1}^Ny_i^E(t)-\frac{(1-f_I)}{f_I}Y_{EE}(t)\approx\\
&\approx\bigg(\frac{1}{f_I}-\frac{(1-f_I)}{f_I}\bigg)\frac{1}{N}\sum_{i=1}^{N}y_i^{E}(t)=\\
&=\frac{1}{N}\sum_{i=1}^{N}y_i^{E}(t).
\end{split}
\label{ApproxYei}
\end{equation}
Finally, we can estimate $Y_E$ as:
\begin{equation}
\begin{split}
Y_E(t)&=(1-f_I)Y_{EE}(t)-f_IY_{EI}(t)=\\
&=(1-2f_I)\frac{1}{N}\sum_{i=1}^Ny_i^E(t).
\end{split}
\label{ApproxYe}
\end{equation}
Remark that the above expression for $Y_E$ is obtained without grouping the neurons in excitatory and inhibitory classes, but provided $f_I$, the fraction of inhibitory neurons, is eventually known. As we lack information on the corresponding field $Y_I$, we can run the reconstruction scheme in a simplified setting which is solely targeted to reconstructing the probability distributions $P(a)$ and $P_E(\tilde{k})$. In Fig. \ref{Pannello1}D, the results of the revisited inversion method are displayed, having set $f_I$ to the correct value, i.e. assuming the relative proportion of excitatory and inhibitory neurons which has been effectively employed in generating the synthetic dataset. The reconstruction algorithm is still capable of returning a faithful representation of both $P(a)$ and $P_E(\tilde{k})$. Conversely, when $f_I$ is set to zero, the reconstruction scheme compensates for the missing inhibitory component by predicting a reduced average connectivity of the excitatory population, as compared to the correct value assumed in the data generation scheme, see Fig. \ref{Pannello1}E.
In the following, we will apply the reconstruction scheme in this latter version to the analysis of 2PLS microscope images of living zebrafish larva.

\begin{figure*}[!ht]
\centering
\includegraphics[width=0.85\textwidth]{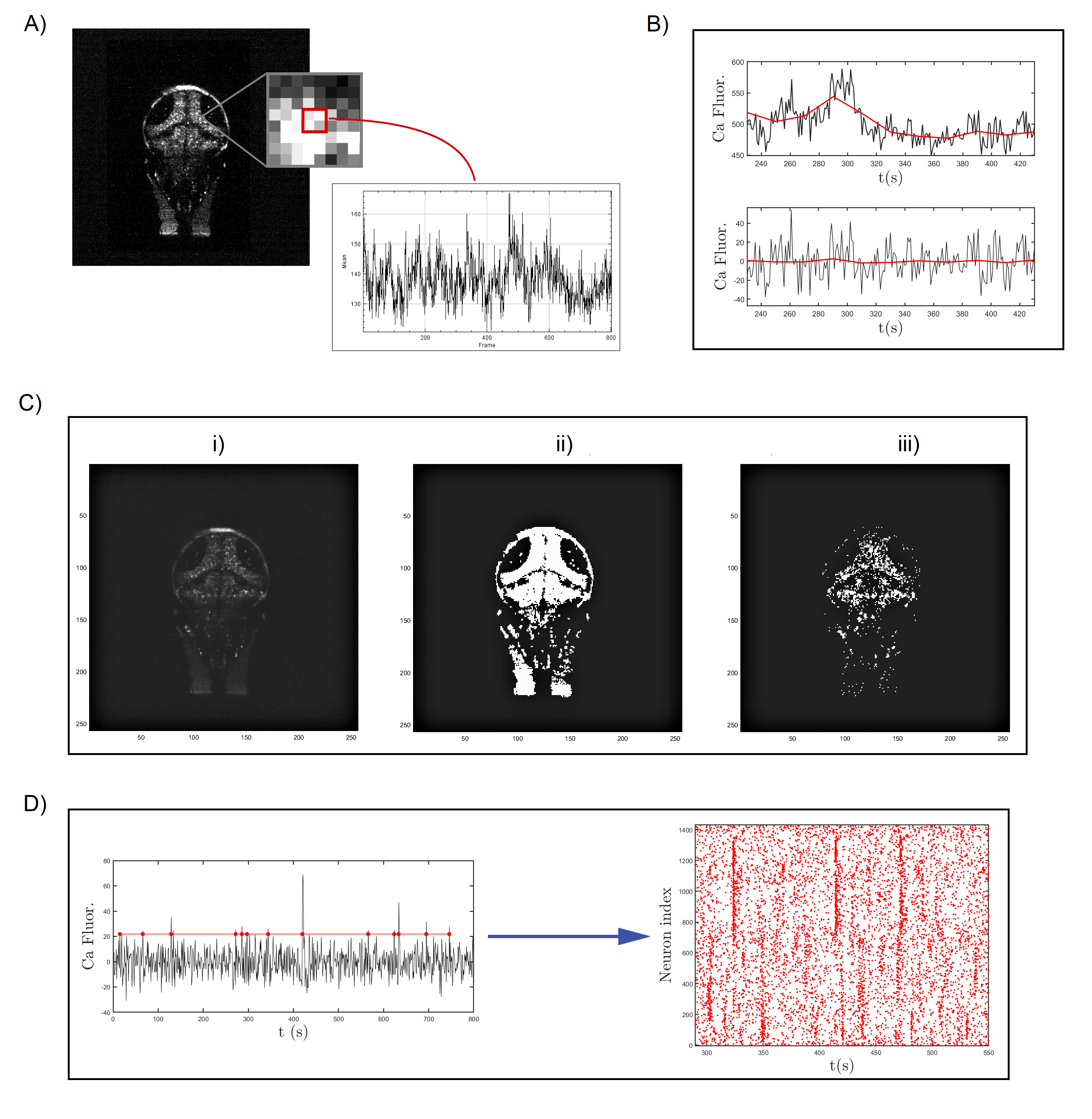}
\caption{\footnotesize Main steps of experimental data elaboration. Every layer of the imaged 3D zebrafish brain is spatially downsampled, as shown in panel A) in order to obtain signals from pixel ensembles of size comparable to a neuron ($2\times 2$ pixels). B) Detrending for slow oscillations by subtraction of moving average. C) Results of neurons selection for one of the layers: (i) raw data, (ii) selection of pixels with maximum value above a fixed threshold, and (iii) only pixels with skeweness larger than 0.4. At the end of the neurons selection procedure the number of identified neurons in the whole brain is about $5\cdot 10^4$. D) Procedure to obtain the experimental raster plot starting from calcium fluorescence time series of the selected neurons. A spike is identified by its upwards threshold crossing time.}
\label{Pannello2}
\end{figure*}

\subsection{Experimental data}
\label{Sec_data}

In this section we apply our reconstruction framework to calcium fluorescence microscopy data of zebrafish larva brain. {Indeed, each time a neuron fires an action potential, the strong depolarization occurring triggers a rapid and transient increase of intracellular calcium concentration \cite{baker1971depolarization}. Thus, following calcium-dependent fluorescence dynamics represents an indirect measurement of neuronal spiking activity \cite{grienberger2012imaging}.}
A description of the experimental set up is provided in the Methods section.

\subsubsection{Data processing}

In order to apply the inverse scheme to real data, it is necessary to pre-process the wide field calcium images  with the purpose of first identifying the location of the neurons.  From individual traces of each spotted neuron, we will single out the spiking events, record the times of occurrence and build up the corresponding raster plot. This will serve as input to the reconstruction algorithm. 

To this aim, data are first downsampled to $2\times2$ pixels so that the new pixel size is comparable to the neurons' nuclear size (Fig. \ref{Pannello2}A).
For every new pixel, the maximum value of the calcium fluorescence is calculated and only the pixels with maximum intensity projection (MIP) above a threshold are identified as neurons.
The threshold is fixed to the average MIP value over all the image pixels. 
Furthermore, we operate a moving average to remove low frequency fluctuation (Fig. \ref{Pannello2}B) and  select as presumed neurons the pixels that show large asymmetry in the recorded traces. 
To implement this step, we compute the \textit{skewness} from individual time series of the calcium activity and classify as neurons the pixels that yield a sufficiently skewed signal\footnote{\footnotesize The skewness threshold is here put to 0.4. In doing so we select a number of putative neurons which is comparable to the known size of a zebrafish larva brain \cite{naumann2010monitoring}}. 
Figure \ref{Pannello2}C outlines the different phases of the process for one of the layers of the collection. More specifically, in Fig. \ref{Pannello2}C.i the MIP of the down-sampled pixels are depicted, in grey scale. In Fig. \ref{Pannello2}C.ii a binary representation of the whole brain is displayed where only pixels with MIP above the fixed threshold are highlighted. Lastly, in Fig. \ref{Pannello2}C.iii only the pixels which exhibit a strong asymmetric signal, i.e. the neurons with skewness above the imposed threshold, are shown. At the end of the selection process the number of identified neurons is around $(1\div 3)\times10^3$ for each layer, which correspond to a total of $49\times10^3$.

Once the neurons are identified, we proceed to construct the raster plot. To this aim, for each selected neuron we analyze the time series of the calcium fluorescence to remove the background noise and detect events, which we call spikes.
More specifically, a spike is defined as the time of threshold crossing. The thresholds are set to the mean value of the recorded time series plus two times the associated standard deviation.
In addition, in order to avoid double detections due to noise, we discarded all events that succeeded the previous event by less than a minimum inter-event interval of 5 data points (5 seconds)\footnote{An additional analysis has been carried out using a minimum inter-event interval of 2 data points. From the corresponding raster plot, the global fields have been calculated and they appear indistinguishable from the ones obtained using a minimum inter-event interval of 5 data points.} \cite{Chen2013}.
The general overview of the spike trains emitted by neurons in a sample layer results in a raster plot (Fig. \ref{Pannello2}D). Time is on the horizontal axis, whereas the vertical axis displays the neuron indices. Each spike of neuron $i$ is associated with a red dot in the row $i$, at the corresponding time of spiking.

\subsubsection{Results}

\begin{figure*}[h]
	\centering
	\includegraphics[width=1\linewidth]{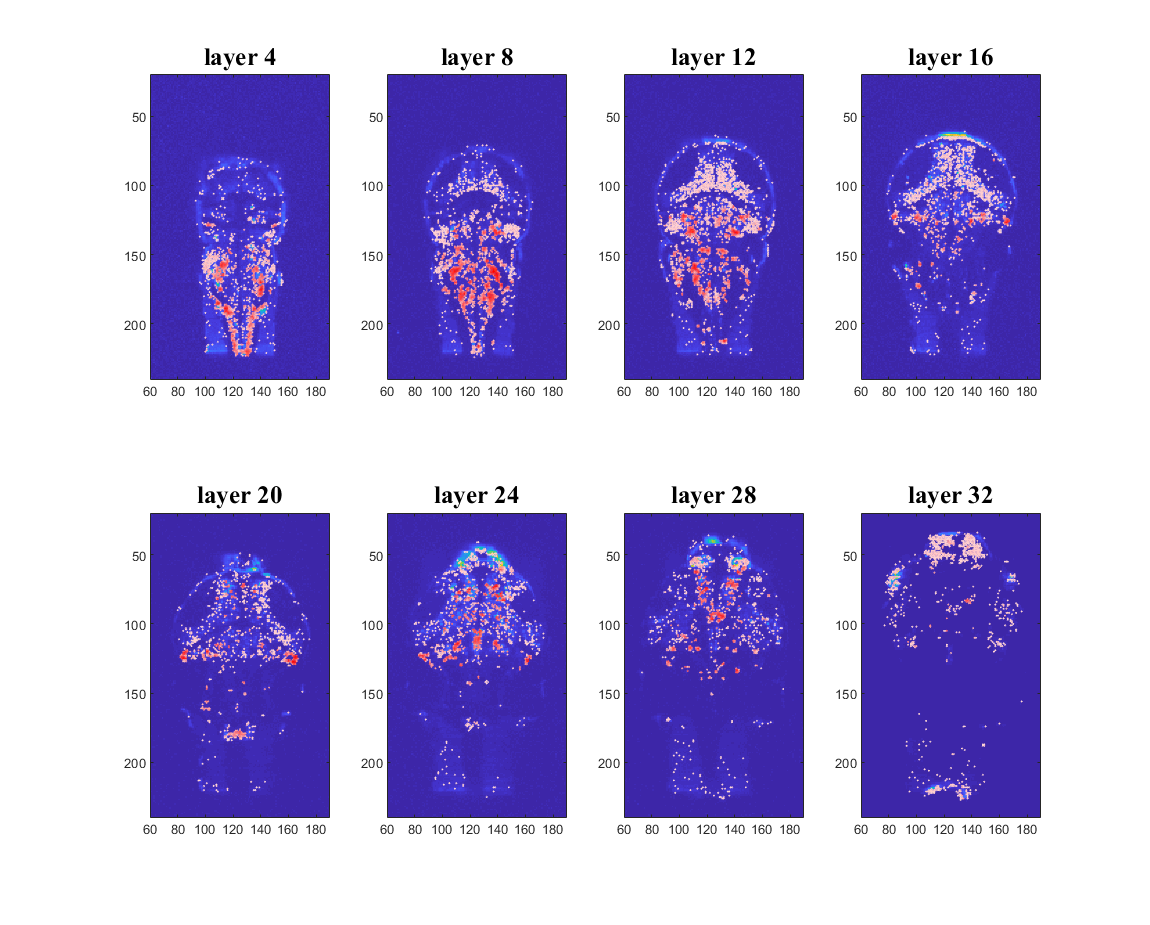}
	\caption{\footnotesize Detected neurons for eight different layers of the zebrafish brain. Colours represent the average cross-correlation of each neuron with all the others selected neurons of the brain.}
	\label{Slices}
\end{figure*}

As described in the previous section, we process 3D calcium fluorescence data so as to identify pixels containing neurons.
Figure \ref{Slices} shows the results of this identification for eight different layers of the zebrafish brain. Colors reflect the average cross-correlation\footnote{Cross-correlation measures the similarity between two series at different time shifts, or lags. In formulae, the cross-correlation between two vectors $x_t$ and $y_t$ at lag $\tau$ is defined as the expected value of the product of the shifted copy of $x_t$ and the complex conjugate of $y_t$, i.e., $R_\tau(x,y)=\mathbf{E}(x_{t+\tau}y_t^*)$, where the asterisk denotes the complex conjugation.} at lag zero of each neuron with all other selected neurons of the brain. The higher the correlation, the more reddish the color displayed. The patterns of correlations are rather symmetric, an observation which can be interpreted as an a posteriori validation of the implemented procedure for automatic neurons selection. A movie which allows to navigate across successive layers of the whole 3D stack can be found in the SM. 

The processing of data explained above allows us to obtain an experimental raster plot describing the events, or spikes, associated to each neuron. Indeed, the raster plot contains information about the spike train function $S_i(t)$, for all neurons $i$, and can be readily employed to recover the global field $Y_E$ to be supplied as an input to the reconstruction scheme. More explicitly, the experimentally determined $S_i(t)$ is used to integrate Eqs. (\ref{Nuerotrans}) and (\ref{NuerotransFraction}), by breaking the coupling with Eq. (\ref{MemPot}) which sets the evolution of the membrane potential\footnote{The reactions parameters are set to the nominal values as declared above \cite{diVolo3}. The same parameters are assumed in the reconstruction, and, in this respect, the model equations acts as a filter to transform the supplied fluorescence data in the ideal input for the inverse procedure to run.}. This is of great advantage since Eq. (\ref{MemPot}) contains the specific information about the network connections, i.e., the adjacency matrix elements, which are a priori unknown. In other words, the raster plot provides us a way to compute the global field (input of the reconstruction process described in section \ref{ReconstrucionScheme}) without knowing the underlying structure of the network.

\begin{figure*}[!ht]
\centering
\includegraphics[width=0.85\textwidth]{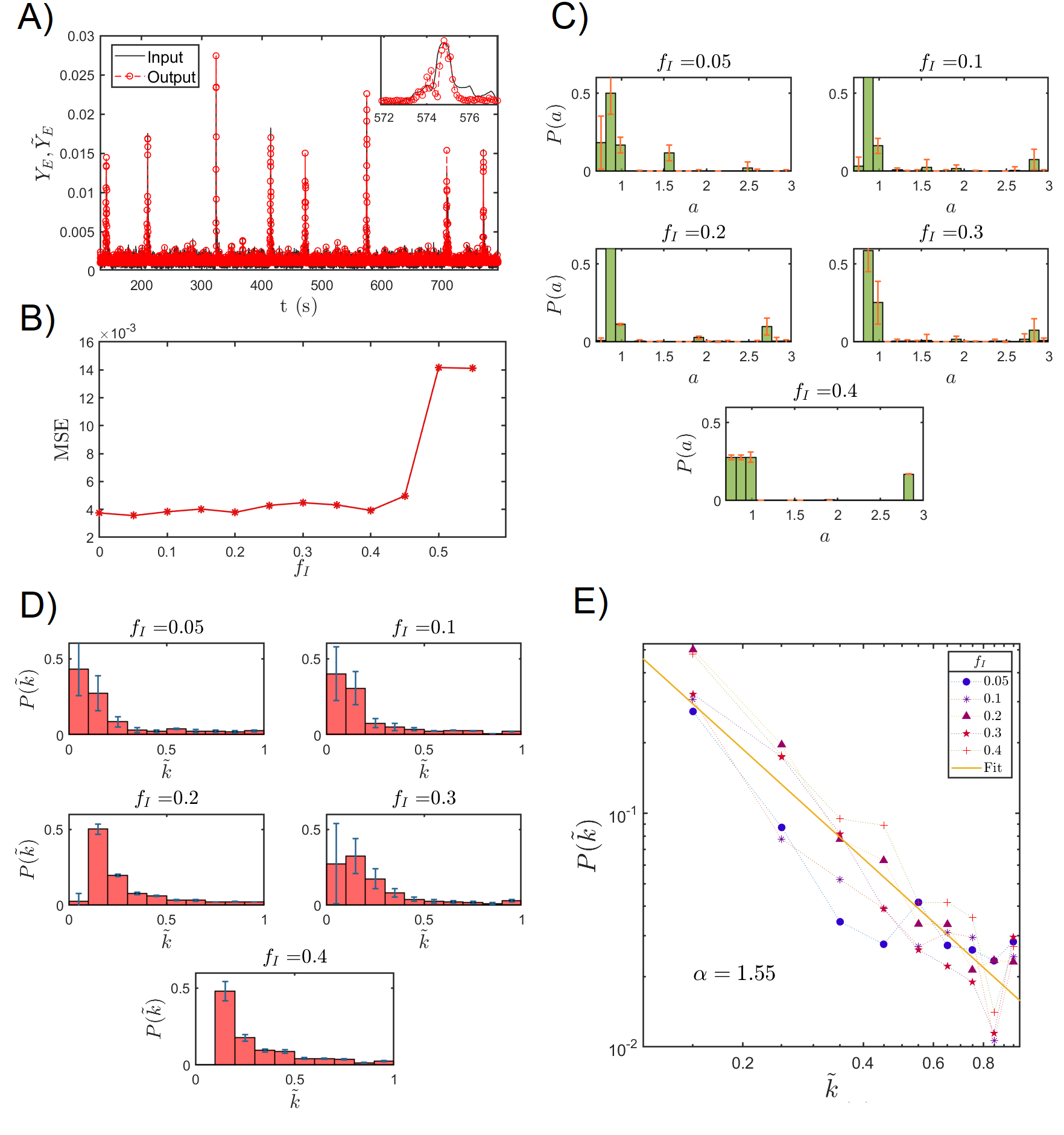}
\caption{\footnotesize Reconstruction of the two distributions $P(a)$ and $P_E(\tilde{k})$ from calcium fluorescence microscopy data of zebrafish larva brain. A) Comparison between the global field obtained from the experimental raster plot and the one that follows the reconstructed distributions $P_E(\tilde{k})$ and $P(a)$, for a fraction $f_I=0.05$ of inhibitory neuron. B) Mean square error $MSE=1/T\sum_{t}^{T}(Y(t)-\tilde{Y}(t))^2$ for different choices of $f_I$. C-D) Reconstructed probability distributions $P(a)$ and $P(\tilde{k})$ for five different choices of $f_I$. For each setting, we ran $50$ independent realizations of the iterative reconstruction scheme, starting from different initial conditions. The histograms represent the mean values and the error bars stands for the associated variance. E) In-degree probability distributions in logarithmic scale and their best linear fitting.}
\label{Pannello3}
\end{figure*}

Since the true fraction $f_I$ of inhibitory neurons is unknown, the global field $Y_E$ in the approximated form [Eq. (\ref{ApproxYe})] is computed for different values of $f_I$.
In particular, we first reconstruct the in-degree distribution $P_E(\tilde{k})$ for the excitatory neurons and the external current distribution $P(a)$ for different values of $f_I$ and we store the results\footnote{The coupling strength $g$ is set to the (experimentally justified) value of 30 \cite{diVolo3,TUM,volman} adopted in the forward simulations of the model.}.
Secondly, we computed the reconstruction error $MSE=1/T\sum_{t}^{T}(Y(t)-\tilde{Y}(t))^2$ for all the considered values of $f_I$, comparing the estimated field $\tilde{Y}_E$ with the one used as input in the reconstruction scheme.
Figure \ref{Pannello3}A reports on this comparison in the case of $f_I=0.05$. In Fig. \ref{Pannello3}B, the $MSE$ is plotted against different values of $f_I$. 
Small errors are found over a large (and biologically meaningful) interval of values for $f_I$, approximately $f_I\in(0,0.4]$. 
For this reason, we focus on five different choices of $f_I$, i.e., $f_I=0.05,0.1,0.2,0.3,0.4$, to explore a wide range of results for the reconstruction scheme, when sampling the  region of parameters in which the interpolation of the experimental time series proves accurate. 
The reconstructed distributions $P(a)$ and $P_E(\tilde{k})$ are plotted in Figs. \ref{Pannello3}C,D. For every choice of $f_I$, over the spanned interval, the reconstructed distributions show common features. In particular, the external current distribution $P(a)$ is peaked in the vicinity of the critical value $a=1$ (Fig. \ref{Pannello3}C). The neurons associated to $a>1$ get self-excited and promote the activation of other neurons which would be instead quiescent in the uncoupled limit. The small bumps that are found for relatively large values of the intensities $a$ can be traced back to the high frequency component of the signal to be interpolated, and, as such, bear limited fundamental interests. The large-scale dynamics of the recorded time-series, including the heterogeneous modulation of the macroscopic field oscillations, is instead encoded in the distribution of intensities that define the bulk of $P(a)$, i.e. the limited excerpt of curve which is found in correspondence of the leftmost portion of the support in $a$. 

The reconstructed in-degree distributions $P_E(\tilde{k})$ for the excitatory neurons, at different values of $f_I$, are depicted in Fig. \ref{Pannello3}E, in log-log scale. Although over a limited support in $\tilde{k}$, the obtained distributions seem to display a power law decay, $P_E(\tilde{k})\propto \tilde{k}^{-\alpha}$, the characteristic exponent $\alpha$ being only modestly influenced by the chosen value of $f_I$. Our findings suggest that excitatory neurons are organized in a network with few hubs and many more peripheral nodes. 

\begin{figure*}[!ht]
	\centering
	\includegraphics[width=1\textwidth]{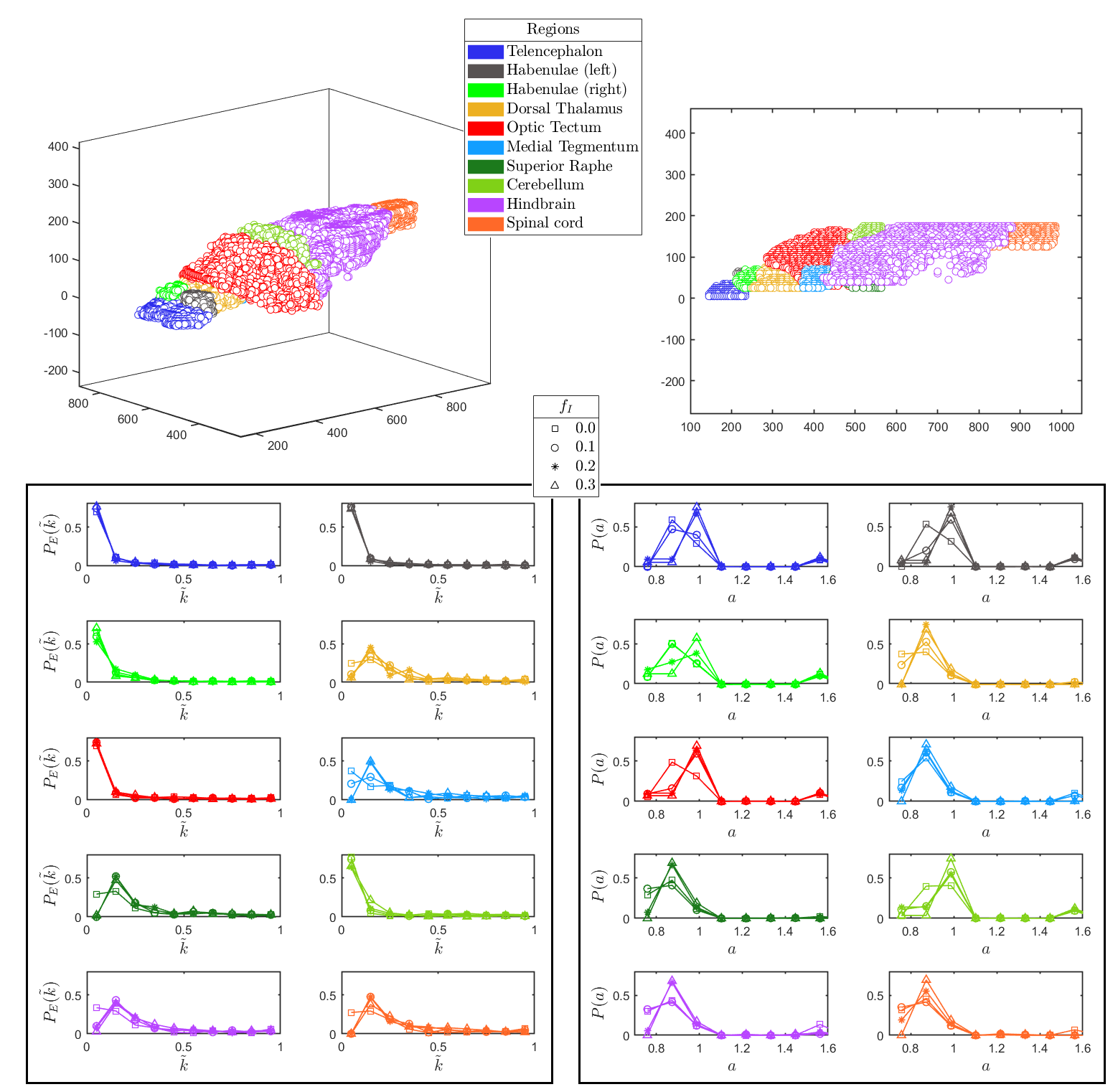}
	\caption{\footnotesize The reconstructed probability distributions $P_E(\tilde{k})$ (left) and $P(a)$ (right) are shown for ten different regions of the brain. The 3D images at the top display the relative spatial positions of the ten selected regions (as listed in the legend).}
	\label{pk_regions50}
\end{figure*}

As an additional test, we partition the full set of identified nodes in $10$ different populations, reflecting distinct anatomical regions, as follows available atlases \cite{naumann2010monitoring,kunst2019cellular}. The reconstruction algorithm is then applied to each of the selected region, treated as independent from the surrounding context, so as to access the local degree distribution.  Results, displayed in Fig. \ref{pk_regions50}, are in line with those reported in  Fig. \ref{Pannello3}. Moreover, neurons characterized by a significant connectivity, the above referenced hubs, seem unevenly distributed across different anatomical regions. 

\section{Conclusions}

Reconstructing structural and functional information from  brain activity represents a
topic of outstanding importance, which can in principle trigger applied and fundamental fallout. In \cite{adam_HMF} we presented, and successfully tested, an inverse scheme which aimed at inferring the distributions of both firing rates and networks connectivity, from global activity fields. The method builds on the Leaky-Integrate and Fire (LIF) model which we modified by the inclusion of quenched disorder, in the assigned individual currents. The imposed degree of heterogeneity in the currents yields non trivial a-periodic patterns, which resemble those recorded in vivo. The dynamical model considered in \cite{adam_HMF} solely included the population of excitatory neurons. Starting from these background, we here have generalized the reconstruction procedure of \cite{adam_HMF} so as to account for the simultaneous presence of both excitatory and inhibitory neurons, while still dealing with the effect of the current heterogeneity. The dynamics of the examined multi-species LIF model is recast in a simplified framework, by grouping together neurons that belong to the same class (inhibitory vs. excitatory), while sharing the similar currents and in-degree. The output of the reduced model, driven by the excitatory and inhibitory global fields, is self-consistently used to seed an iterative scheme which seeks at fitting the supplied fields, via suitably adjusting the unknown distributions. These latter are the distributions of the incoming degrees for, respectively, excitatory and inhibitory neurons, as well as the distribution of the imposed currents. The method is tested on synthetic data and yields satisfying performances. Having in mind applications to real data, we also dealt with a setting where it is not possible to separate the contribution that pertains to the excitatory component from that stemming from the inhibitory counterpart. In this case, we propose and test a procedure which enables to recover the distribution of incoming degrees for the network of excitatory neurons (assumed predominant), while gauging the role exerted by inhibitors.  
 
The devised protocol is then applied to {whole-brain functional data resulting from} light-sheet {calcium} imaging of a zebrafish larva. The experimental input is processed with an automatic procedure which allows us to identify putative neurons {, and to extract their fluorescence signal. Remarkably, the cross-correlation maps produced show a high grade of clusterization, which faithfully matches the anatomical boundaries of multiple brain regions identified using zebrafish brain atlases \cite{randlett2015whole,kunst2019cellular}.} From the calcium signal displayed by {each} selected neuron we build up an experimental representation of the raster plot of the spiking activity of the zebrafish brain, which forms the input to the reconstruction scheme. A power law distribution of the in-coming connectivity of excitatory neurons is found, which is only modestly affected by the imposed fraction of inhibitors. Local degree distributions are also reconstructed by analysing the signal from specific regions, which correspond to distinct anatomical areas. {Interestingly, the anatomical districts considered in the analysis can be divided into two different groups, according to the reconstructed probability distributions of both their excitatory incoming connections $P_E(\tilde{k})$ and excitability $P(a)$. The first group, including dorsal thalamus, medial tegmentum, superior raphe, hindbrain and spinal cord, is characterized by higher connections and lower excitability. Conversely, the second group, comprising telencephalon, habenulae, optic tectum and cerebellum, is described by lower connections and higher excitability. The reconstruction scheme reflects the specific functional connectivity of the larval brain during spontaneous activity precisely under these experimental conditions. Indeed, during measurements the zebrafish larva is embedded in agarose, and thus exposed to a diffused tactile stimulation, which could explain the higher incoming connections calculated for the dorsal thalamus, a sensory relay station \cite{northcutt2008forebrain,mueller2012thalamus}. Furthermore, despite mechanically and pharmacologically immobilized, larval attempts to escape the restrained condition could account for the higher incoming excitatory connections calculated for dorsal raphe (whose activity has been correlated with arousal state, vigilance and responsiveness \cite{yokogawa2012dorsal}) and for the most caudal regions, namely hindbrain and spinal cord, responsible for the initiation of motor behaviours \cite{garcia2010circuits,kinkhabwala2011structural}. Moreover, in this scenario we observe a lower probability distribution of incoming excitatory connections for cerebellum. This result may be associated to the function of motor coordination and refinement \cite{heap2013cerebellar,kaslin2016zebrafish} of this region, typically related to the actual execution of a movement. Finally, since the measurement is performed in complete darkness, the larva is not exposed to any visual cue (the IR laser used for two-photon excitation is not perceived by the larval visual system) and this could explain the lower incoming connections calculated for the optic tectum, the main retinorecipient structure in the zebrafish brain \cite{sajovic1982visual,gebhardt2019interhemispheric}.}  The large-scale oscillations in the recorded time series reflect back in the recovered distribution of currents: a significant fraction of neurons appear to operate in the quiescent state, while a minority self-excite to orchestrate the dynamics of the ensemble. The existence of possible correlations between individual connectivities and associated neurons’ currents cannot be resolved within the proposed approach and defines an interesting target for future investigations.

\section{Materials and Methods}

\subsection{Validity of the HMF approximation}

We here challenge the predictive ability of the HMF approximation. To this end, we first calculate the average inter-spike interval (ISI) -- the average distance in time between successive spikes -- for the system in its original formulation, i.e. in the space of the nodes. The computed ISI is confronted to the homologous quantity obtained under the HMF scenario. The comparison is drawn in Fig. \ref{HMF}, for both excitatory and inhibitory neurons, and confirm the accuracy of the reduced HMF scheme.

\begin{figure*}[!ht]
	\centering
	\includegraphics[width=0.6\textwidth]{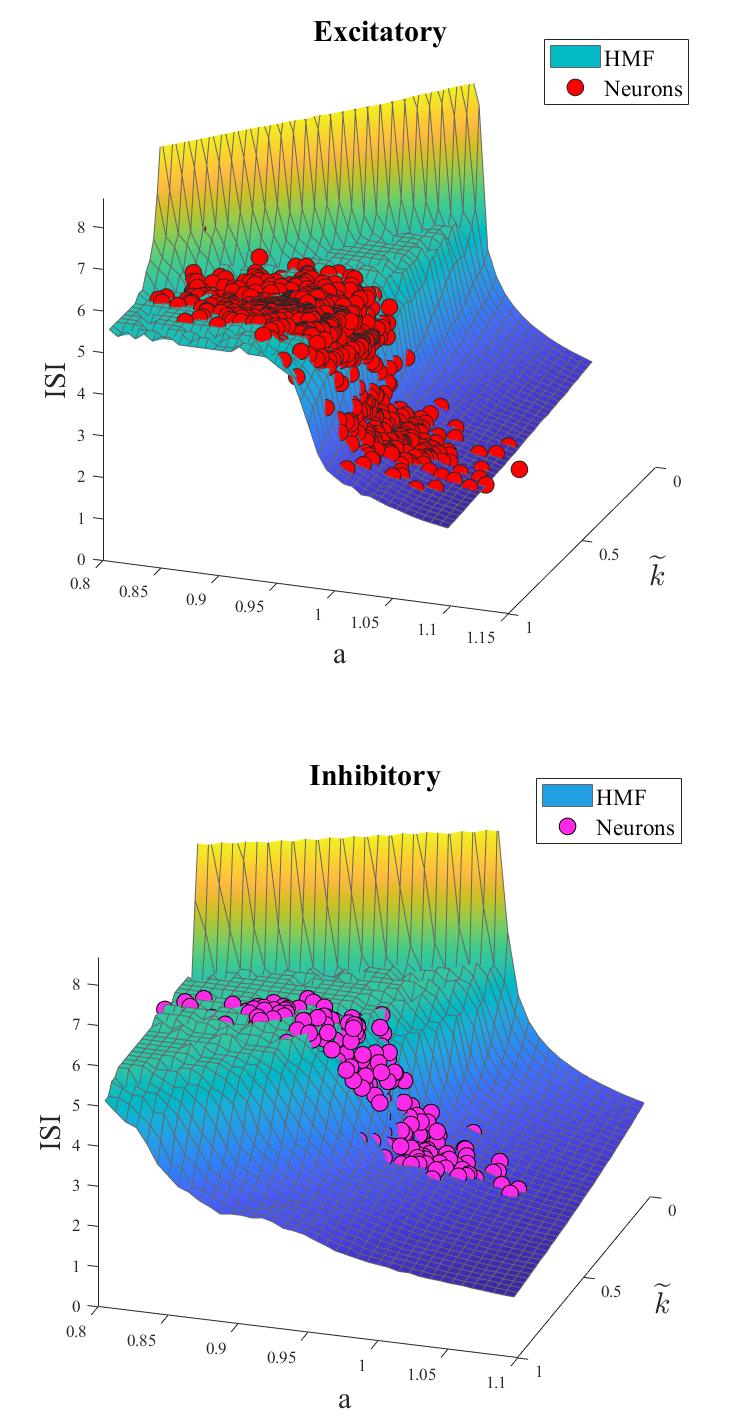}
	\caption{\footnotesize The ISI is computed for both excitatory (red symbols, top panel) and inhibitory (purple symbols, bottom panel) neurons. The prediction based on the HMF approximation yields the continuous curves. Here $N=1000$ and $f_I=0.2$.}
	\label{HMF}
\end{figure*}

\subsection{Experimental setup}

The experimental optical setup employed is a modified version of the setup described in detail in \cite{de2020effects}.
Briefly, $930$ nm NIR light is generated by a pulsed titanium-sapphire oscillator (Chameleon Ultra II,   Coherent) and conditioned by a pulse compressor (PreComp, Coherent). After being attenuated by a combination of a half-wave plate and a Glan–Thompson polarizer, the beam passes through an electro-optical modulator that periodically switches its polarization plane orientation between two states: parallel or orthogonal to the optical table surface. Then a pair of retarders are used to pre-compensate for polarization distortions. After that, the beam is routed to the galvanometric mirror assembly and scanned by a resonant galvanometric mirror (CRS-8 kHz, Cambridge Technology) along larval rostro-caudal direction, to generate the digitally-scanned LS, and by a closed-loop galvanometric mirror (6215H,   Cambridge   Technology) along larval dorso-ventral direction. Finally, the beam is relayed to an objective (XLFLUOR4X/340/0,28,Olympus), placed at the lateral side of the larva, by a scan-lens ($50$ mm focal length), a tube-lens ($75$ mm focal length) and a pair of relay lenses ($250$ mm and $200$ mm focal lengths).

Differently from the setup described in \cite{de2020effects}, a polarizing beam splitter (PBS) is present between the tube-lens and the first relay lens and the optical components downstream the PBS are duplicated on the opposite side of the larva. In this way the excitation light is steered on one optical arm or on the other depending on its instantaneous polarization orientation by the PBS. 

The detection arm of the microscope is identical to what described in \cite{de2020effects}. A water-immersion objective (XLUMPLFLN20XW,  Olympus)  placed on the dorsal side of the larva collects the emitted green fluorescent light while being scanned along the axial dimension by an objective scanner (PIFOC P-725.4CD,Physik Instrumente) synchronously with the closed-loop galvanometric mirror oscillations.
The optical image is then spectrally filtered (FF01-510/84-25 nm BrightLine®,Semrock) and projected on a camera (ORCA-Flash4.0 V3, Hamamatsu) sensor by a sequence of two  tube   lenses   ($300$ mm and $200$ mm   focal   lengths) and an objective (UPLFLN10X2,   Olympus).

The   larvae were   imaged   with   volumetric   acquisitions (frequency: $1$ Hz) composed by $31$ planes spaced by $5$ $\mu m$ and with a pixel size of about $2\times 2$ $\mu m^2$.

We employed a transgenic strain of zebrafish larvae Tg(elavl3:H2B-GCaMP6s) \cite{vladimirov2014light,mullenbroich2018bessel} in   homozygous  albino  background that express the fluorescent calcium indicator GCaMP6s under a pan-neuronal promoter and with a nuclear localization.
 Sample mounting was performed as described in \cite{turrini2017optical}. Briefly, before the acquisition each larva was immersed in a   solution   of the paralyzing agent   d-tubocurarine   ($2$   mM;  93750,   Sigma-Aldrich),   included in $1.5\%$ (w/v) low gelling temperature agarose (A9414, Sigma-Aldrich) in fish water ($150$ mg/L Instant Ocean, $6.9$ mg/L NaH2PO4, $12.5$ mg/L Na2HPO4, pH $7.2$) and mounted on a custom-made glass support immersed in thermostated fish water.
The animals were maintained according to standard procedures \cite{westerfield2000zebrafish} and observed   at   $4$   days   post   fertilization.   Fish maintenance and handling were carried out in accordance with European and Italian law on animal experimentation (D.L. 4 March 2014, no. 26).

\bibliographystyle{apsrev4-1}
\bibliography{bibliografia.bib}

\begin{thebibliography}{43}%
\makeatletter
\providecommand \@ifxundefined [1]{%
 \@ifx{#1\undefined}
}%
\providecommand \@ifnum [1]{%
 \ifnum #1\expandafter \@firstoftwo
 \else \expandafter \@secondoftwo
 \fi
}%
\providecommand \@ifx [1]{%
 \ifx #1\expandafter \@firstoftwo
 \else \expandafter \@secondoftwo
 \fi
}%
\providecommand \natexlab [1]{#1}%
\providecommand \enquote  [1]{``#1''}%
\providecommand \bibnamefont  [1]{#1}%
\providecommand \bibfnamefont [1]{#1}%
\providecommand \citenamefont [1]{#1}%
\providecommand \href@noop [0]{\@secondoftwo}%
\providecommand \href [0]{\begingroup \@sanitize@url \@href}%
\providecommand \@href[1]{\@@startlink{#1}\@@href}%
\providecommand \@@href[1]{\endgroup#1\@@endlink}%
\providecommand \@sanitize@url [0]{\catcode `\\12\catcode `\$12\catcode
  `\&12\catcode `\#12\catcode `\^12\catcode `\_12\catcode `\%12\relax}%
\providecommand \@@startlink[1]{}%
\providecommand \@@endlink[0]{}%
\providecommand \url  [0]{\begingroup\@sanitize@url \@url }%
\providecommand \@url [1]{\endgroup\@href {#1}{\urlprefix }}%
\providecommand \urlprefix  [0]{URL }%
\providecommand \Eprint [0]{\href }%
\providecommand \doibase [0]{http://dx.doi.org/}%
\providecommand \selectlanguage [0]{\@gobble}%
\providecommand \bibinfo  [0]{\@secondoftwo}%
\providecommand \bibfield  [0]{\@secondoftwo}%
\providecommand \translation [1]{[#1]}%
\providecommand \BibitemOpen [0]{}%
\providecommand \bibitemStop [0]{}%
\providecommand \bibitemNoStop [0]{.\EOS\space}%
\providecommand \EOS [0]{\spacefactor3000\relax}%
\providecommand \BibitemShut  [1]{\csname bibitem#1\endcsname}%
\let\auto@bib@innerbib\@empty
\bibitem [{\citenamefont {Schneidman}\ \emph {et~al.}(2006)\citenamefont
  {Schneidman}, \citenamefont {Berry}, \citenamefont {Segev},\ and\
  \citenamefont {Bialek}}]{bialek}%
  \BibitemOpen
  \bibfield  {author} {\bibinfo {author} {\bibfnamefont {E.}~\bibnamefont
  {Schneidman}}, \bibinfo {author} {\bibfnamefont {M.~J.}\ \bibnamefont
  {Berry}}, \bibinfo {author} {\bibfnamefont {R.}~\bibnamefont {Segev}}, \ and\
  \bibinfo {author} {\bibfnamefont {W.}~\bibnamefont {Bialek}},\ }\href@noop {}
  {\bibfield  {journal} {\bibinfo  {journal} {Nature}\ }\textbf {\bibinfo
  {volume} {440}},\ \bibinfo {pages} {1007} (\bibinfo {year}
  {2006})}\BibitemShut {NoStop}%
\bibitem [{\citenamefont {Cocco}\ \emph {et~al.}(2009)\citenamefont {Cocco},
  \citenamefont {Leibler},\ and\ \citenamefont {Monasson}}]{cocco}%
  \BibitemOpen
  \bibfield  {author} {\bibinfo {author} {\bibfnamefont {S.}~\bibnamefont
  {Cocco}}, \bibinfo {author} {\bibfnamefont {S.}~\bibnamefont {Leibler}}, \
  and\ \bibinfo {author} {\bibfnamefont {R.}~\bibnamefont {Monasson}},\
  }\href@noop {} {\bibfield  {journal} {\bibinfo  {journal} {Proceedings of the
  National Academy of Sciences}\ }\textbf {\bibinfo {volume} {106}},\ \bibinfo
  {pages} {14058} (\bibinfo {year} {2009})}\BibitemShut {NoStop}%
\bibitem [{\citenamefont {Dehghani}\ \emph {et~al.}(2016)\citenamefont
  {Dehghani}, \citenamefont {Peyrache}, \citenamefont {Telenczuk},
  \citenamefont {Le~Van~Quyen}, \citenamefont {Halgren}, \citenamefont {Cash},
  \citenamefont {Hatsopoulos},\ and\ \citenamefont {Destexhe}}]{deghani}%
  \BibitemOpen
  \bibfield  {author} {\bibinfo {author} {\bibfnamefont {N.}~\bibnamefont
  {Dehghani}}, \bibinfo {author} {\bibfnamefont {A.}~\bibnamefont {Peyrache}},
  \bibinfo {author} {\bibfnamefont {B.}~\bibnamefont {Telenczuk}}, \bibinfo
  {author} {\bibfnamefont {M.}~\bibnamefont {Le~Van~Quyen}}, \bibinfo {author}
  {\bibfnamefont {E.}~\bibnamefont {Halgren}}, \bibinfo {author} {\bibfnamefont
  {S.~S.}\ \bibnamefont {Cash}}, \bibinfo {author} {\bibfnamefont {N.~G.}\
  \bibnamefont {Hatsopoulos}}, \ and\ \bibinfo {author} {\bibfnamefont
  {A.}~\bibnamefont {Destexhe}},\ }\href {\doibase 10.1038/srep23176}
  {\bibfield  {journal} {\bibinfo  {journal} {Scientific Reports}\ }\textbf
  {\bibinfo {volume} {6}},\ \bibinfo {pages} {23176} (\bibinfo {year}
  {2016})}\BibitemShut {NoStop}%
\bibitem [{\citenamefont {Kaczmarek}\ and\ \citenamefont
  {Levitan}(1987)}]{neuromod}%
  \BibitemOpen
  \bibfield  {author} {\bibinfo {author} {\bibfnamefont {L.~K.}\ \bibnamefont
  {Kaczmarek}}\ and\ \bibinfo {author} {\bibfnamefont {I.~B.}\ \bibnamefont
  {Levitan}},\ }\href@noop {} {\emph {\bibinfo {title} {Neuromodulation: the
  biochemical control of neuronal excitability}}}\ (\bibinfo  {publisher}
  {Oxford University Press},\ \bibinfo {year} {1987})\BibitemShut {NoStop}%
\bibitem [{\citenamefont {Adam}\ \emph
  {et~al.}(2019{\natexlab{a}})\citenamefont {Adam}, \citenamefont {Cecchini},
  \citenamefont {Fanelli}, \citenamefont {Kreuz}, \citenamefont {Livi},
  \citenamefont {di~Volo}, \citenamefont {Mascaro}, \citenamefont {Conti},
  \citenamefont {Scaglione}, \citenamefont {Silvestri} \emph
  {et~al.}}]{adam_HMF}%
  \BibitemOpen
  \bibfield  {author} {\bibinfo {author} {\bibfnamefont {I.}~\bibnamefont
  {Adam}}, \bibinfo {author} {\bibfnamefont {G.}~\bibnamefont {Cecchini}},
  \bibinfo {author} {\bibfnamefont {D.}~\bibnamefont {Fanelli}}, \bibinfo
  {author} {\bibfnamefont {T.}~\bibnamefont {Kreuz}}, \bibinfo {author}
  {\bibfnamefont {R.}~\bibnamefont {Livi}}, \bibinfo {author} {\bibfnamefont
  {M.}~\bibnamefont {di~Volo}}, \bibinfo {author} {\bibfnamefont {A.~L.~A.}\
  \bibnamefont {Mascaro}}, \bibinfo {author} {\bibfnamefont {E.}~\bibnamefont
  {Conti}}, \bibinfo {author} {\bibfnamefont {A.}~\bibnamefont {Scaglione}},
  \bibinfo {author} {\bibfnamefont {L.}~\bibnamefont {Silvestri}},  \emph
  {et~al.},\ }\href@noop {} {\bibfield  {journal} {\bibinfo  {journal} {arXiv
  preprint arXiv:1910.05761}\ } (\bibinfo {year}
  {2019}{\natexlab{a}})}\BibitemShut {NoStop}%
\bibitem [{\citenamefont {Barrat}\ \emph {et~al.}(2008)\citenamefont {Barrat},
  \citenamefont {Barthelemy},\ and\ \citenamefont {Vespignani}}]{barrat2008}%
  \BibitemOpen
  \bibfield  {author} {\bibinfo {author} {\bibfnamefont {A.}~\bibnamefont
  {Barrat}}, \bibinfo {author} {\bibfnamefont {M.}~\bibnamefont {Barthelemy}},
  \ and\ \bibinfo {author} {\bibfnamefont {A.}~\bibnamefont {Vespignani}},\
  }\href@noop {} {\emph {\bibinfo {title} {Dynamical processes on complex
  networks}}}\ (\bibinfo  {publisher} {Cambridge university press},\ \bibinfo
  {year} {2008})\BibitemShut {NoStop}%
\bibitem [{\citenamefont {Dorogovtsev}\ \emph {et~al.}(2008)\citenamefont
  {Dorogovtsev}, \citenamefont {Goltsev},\ and\ \citenamefont
  {Mendes}}]{dorogovtsev2008}%
  \BibitemOpen
  \bibfield  {author} {\bibinfo {author} {\bibfnamefont {S.~N.}\ \bibnamefont
  {Dorogovtsev}}, \bibinfo {author} {\bibfnamefont {A.~V.}\ \bibnamefont
  {Goltsev}}, \ and\ \bibinfo {author} {\bibfnamefont {J.~F.}\ \bibnamefont
  {Mendes}},\ }\href@noop {} {\bibfield  {journal} {\bibinfo  {journal}
  {Reviews of Modern Physics}\ }\textbf {\bibinfo {volume} {80}},\ \bibinfo
  {pages} {1275} (\bibinfo {year} {2008})}\BibitemShut {NoStop}%
\bibitem [{\citenamefont {Pastor-Satorras}\ and\ \citenamefont
  {Vespignani}(2001)}]{pastor2001}%
  \BibitemOpen
  \bibfield  {author} {\bibinfo {author} {\bibfnamefont {R.}~\bibnamefont
  {Pastor-Satorras}}\ and\ \bibinfo {author} {\bibfnamefont {A.}~\bibnamefont
  {Vespignani}},\ }\href@noop {} {\bibfield  {journal} {\bibinfo  {journal}
  {Physical review letters}\ }\textbf {\bibinfo {volume} {86}},\ \bibinfo
  {pages} {3200} (\bibinfo {year} {2001})}\BibitemShut {NoStop}%
\bibitem [{\citenamefont {Vespignani}(2012)}]{vespignani2012}%
  \BibitemOpen
  \bibfield  {author} {\bibinfo {author} {\bibfnamefont {A.}~\bibnamefont
  {Vespignani}},\ }\href@noop {} {\bibfield  {journal} {\bibinfo  {journal}
  {Nature physics}\ }\textbf {\bibinfo {volume} {8}},\ \bibinfo {pages} {32}
  (\bibinfo {year} {2012})}\BibitemShut {NoStop}%
\bibitem [{\citenamefont {Burioni}\ \emph {et~al.}(2014)\citenamefont
  {Burioni}, \citenamefont {Casartelli}, \citenamefont {di~Volo}, \citenamefont
  {Livi},\ and\ \citenamefont {Vezzani}}]{diVolo1}%
  \BibitemOpen
  \bibfield  {author} {\bibinfo {author} {\bibfnamefont {R.}~\bibnamefont
  {Burioni}}, \bibinfo {author} {\bibfnamefont {M.}~\bibnamefont {Casartelli}},
  \bibinfo {author} {\bibfnamefont {M.}~\bibnamefont {di~Volo}}, \bibinfo
  {author} {\bibfnamefont {R.}~\bibnamefont {Livi}}, \ and\ \bibinfo {author}
  {\bibfnamefont {A.}~\bibnamefont {Vezzani}},\ }\href {\doibase
  10.1038/srep04336} {\bibfield  {journal} {\bibinfo  {journal} {Scientific
  Reports}\ }\textbf {\bibinfo {volume} {4}},\ \bibinfo {pages} {4336}
  (\bibinfo {year} {2014})}\BibitemShut {NoStop}%
\bibitem [{\citenamefont {di~Volo}\ \emph {et~al.}(2014)\citenamefont
  {di~Volo}, \citenamefont {Burioni}, \citenamefont {Casartelli}, \citenamefont
  {Livi},\ and\ \citenamefont {Vezzani}}]{diVolo2}%
  \BibitemOpen
  \bibfield  {author} {\bibinfo {author} {\bibfnamefont {M.}~\bibnamefont
  {di~Volo}}, \bibinfo {author} {\bibfnamefont {R.}~\bibnamefont {Burioni}},
  \bibinfo {author} {\bibfnamefont {M.}~\bibnamefont {Casartelli}}, \bibinfo
  {author} {\bibfnamefont {R.}~\bibnamefont {Livi}}, \ and\ \bibinfo {author}
  {\bibfnamefont {A.}~\bibnamefont {Vezzani}},\ }\href {\doibase
  10.1103/physreve.90.022811} {\bibfield  {journal} {\bibinfo  {journal}
  {Physical Review. E, statistical, nonlinear, and soft matter physics}\
  }\textbf {\bibinfo {volume} {90}},\ \bibinfo {pages} {022811} (\bibinfo
  {year} {2014})}\BibitemShut {NoStop}%
\bibitem [{\citenamefont {di~Volo}\ \emph {et~al.}(2016)\citenamefont
  {di~Volo}, \citenamefont {Burioni}, \citenamefont {Casartelli}, \citenamefont
  {Livi},\ and\ \citenamefont {Vezzani}}]{diVolo3}%
  \BibitemOpen
  \bibfield  {author} {\bibinfo {author} {\bibfnamefont {M.}~\bibnamefont
  {di~Volo}}, \bibinfo {author} {\bibfnamefont {R.}~\bibnamefont {Burioni}},
  \bibinfo {author} {\bibfnamefont {M.}~\bibnamefont {Casartelli}}, \bibinfo
  {author} {\bibfnamefont {R.}~\bibnamefont {Livi}}, \ and\ \bibinfo {author}
  {\bibfnamefont {A.}~\bibnamefont {Vezzani}},\ }\href@noop {} {\bibfield
  {journal} {\bibinfo  {journal} {Physical Review E}\ }\textbf {\bibinfo
  {volume} {93}},\ \bibinfo {pages} {012305} (\bibinfo {year}
  {2016})}\BibitemShut {NoStop}%
\bibitem [{\citenamefont {Adam}\ \emph
  {et~al.}(2019{\natexlab{b}})\citenamefont {Adam}, \citenamefont {Fanelli},
  \citenamefont {Carletti},\ and\ \citenamefont {Innocenti}}]{AFCI}%
  \BibitemOpen
  \bibfield  {author} {\bibinfo {author} {\bibfnamefont {I.}~\bibnamefont
  {Adam}}, \bibinfo {author} {\bibfnamefont {D.}~\bibnamefont {Fanelli}},
  \bibinfo {author} {\bibfnamefont {T.}~\bibnamefont {Carletti}}, \ and\
  \bibinfo {author} {\bibfnamefont {G.}~\bibnamefont {Innocenti}},\ }\href
  {\doibase 10.1140/epjb/e2019-90700-3} {\bibfield  {journal} {\bibinfo
  {journal} {The European Physical Journal B}\ }\textbf {\bibinfo {volume}
  {92}},\ \bibinfo {pages} {99} (\bibinfo {year}
  {2019}{\natexlab{b}})}\BibitemShut {NoStop}%
\bibitem [{\citenamefont {Schiene}\ \emph {et~al.}(1996)\citenamefont
  {Schiene}, \citenamefont {Bruehl}, \citenamefont {Zilles}, \citenamefont
  {Qu}, \citenamefont {Hagemann}, \citenamefont {Kraemer},\ and\ \citenamefont
  {Witte}}]{HyperExcitability1}%
  \BibitemOpen
  \bibfield  {author} {\bibinfo {author} {\bibfnamefont {K.}~\bibnamefont
  {Schiene}}, \bibinfo {author} {\bibfnamefont {C.}~\bibnamefont {Bruehl}},
  \bibinfo {author} {\bibfnamefont {K.}~\bibnamefont {Zilles}}, \bibinfo
  {author} {\bibfnamefont {M.}~\bibnamefont {Qu}}, \bibinfo {author}
  {\bibfnamefont {G.}~\bibnamefont {Hagemann}}, \bibinfo {author}
  {\bibfnamefont {M.}~\bibnamefont {Kraemer}}, \ and\ \bibinfo {author}
  {\bibfnamefont {O.~W.}\ \bibnamefont {Witte}},\ }\href {\doibase
  10.1097/00004647-199609000-00014} {\bibfield  {journal} {\bibinfo  {journal}
  {Journal of Cerebral Blood Flow \& Metabolism}\ }\textbf {\bibinfo {volume}
  {16}},\ \bibinfo {pages} {906} (\bibinfo {year} {1996})}\BibitemShut
  {NoStop}%
\bibitem [{\citenamefont {Neumann-Haefelin}\ \emph {et~al.}(1995)\citenamefont
  {Neumann-Haefelin}, \citenamefont {Hagemann},\ and\ \citenamefont
  {Witte}}]{HyperExcitability2}%
  \BibitemOpen
  \bibfield  {author} {\bibinfo {author} {\bibfnamefont {T.}~\bibnamefont
  {Neumann-Haefelin}}, \bibinfo {author} {\bibfnamefont {G.}~\bibnamefont
  {Hagemann}}, \ and\ \bibinfo {author} {\bibfnamefont {O.}~\bibnamefont
  {Witte}},\ }\href {\doibase 10.1016/0304-3940(95)11677-o} {\bibfield
  {journal} {\bibinfo  {journal} {Neuroscience Letters}\ }\textbf {\bibinfo
  {volume} {193}},\ \bibinfo {pages} {101—104} (\bibinfo {year}
  {1995})}\BibitemShut {NoStop}%
\bibitem [{\citenamefont {Berger}\ \emph {et~al.}(2019)\citenamefont {Berger},
  \citenamefont {Varriale}, \citenamefont {van Kessenich}, \citenamefont
  {Herrmann},\ and\ \citenamefont {de~Arcangelis}}]{HyperExcitability3}%
  \BibitemOpen
  \bibfield  {author} {\bibinfo {author} {\bibfnamefont {D.}~\bibnamefont
  {Berger}}, \bibinfo {author} {\bibfnamefont {E.}~\bibnamefont {Varriale}},
  \bibinfo {author} {\bibfnamefont {L.~M.}\ \bibnamefont {van Kessenich}},
  \bibinfo {author} {\bibfnamefont {H.~J.}\ \bibnamefont {Herrmann}}, \ and\
  \bibinfo {author} {\bibfnamefont {L.}~\bibnamefont {de~Arcangelis}},\ }\href
  {\doibase 10.1038/s41598-019-50946-y} {\bibfield  {journal} {\bibinfo
  {journal} {Scientific Reports}\ }\textbf {\bibinfo {volume} {9}},\ \bibinfo
  {pages} {15858} (\bibinfo {year} {2019})}\BibitemShut {NoStop}%
\bibitem [{\citenamefont {Wolf}\ \emph {et~al.}(2015)\citenamefont {Wolf},
  \citenamefont {Supatto}, \citenamefont {Debr{\'e}geas}, \citenamefont
  {Mahou}, \citenamefont {Kruglik}, \citenamefont {Sintes}, \citenamefont
  {Beaurepaire},\ and\ \citenamefont {Candelier}}]{wolf2015whole}%
  \BibitemOpen
  \bibfield  {author} {\bibinfo {author} {\bibfnamefont {S.}~\bibnamefont
  {Wolf}}, \bibinfo {author} {\bibfnamefont {W.}~\bibnamefont {Supatto}},
  \bibinfo {author} {\bibfnamefont {G.}~\bibnamefont {Debr{\'e}geas}}, \bibinfo
  {author} {\bibfnamefont {P.}~\bibnamefont {Mahou}}, \bibinfo {author}
  {\bibfnamefont {S.~G.}\ \bibnamefont {Kruglik}}, \bibinfo {author}
  {\bibfnamefont {J.-M.}\ \bibnamefont {Sintes}}, \bibinfo {author}
  {\bibfnamefont {E.}~\bibnamefont {Beaurepaire}}, \ and\ \bibinfo {author}
  {\bibfnamefont {R.}~\bibnamefont {Candelier}},\ }\href@noop {} {\bibfield
  {journal} {\bibinfo  {journal} {Nature Methods}\ }\textbf {\bibinfo {volume}
  {12}},\ \bibinfo {pages} {379} (\bibinfo {year} {2015})}\BibitemShut
  {NoStop}%
\bibitem [{\citenamefont {de~Vito}\ \emph
  {et~al.}(2020{\natexlab{a}})\citenamefont {de~Vito}, \citenamefont
  {Fornetto}, \citenamefont {Ricci}, \citenamefont {M{\"u}llenbroich},
  \citenamefont {Sancataldo}, \citenamefont {Turrini}, \citenamefont
  {Mazzamuto}, \citenamefont {Tiso}, \citenamefont {Sacconi}, \citenamefont
  {Fanelli} \emph {et~al.}}]{de2020twoa}%
  \BibitemOpen
  \bibfield  {author} {\bibinfo {author} {\bibfnamefont {G.}~\bibnamefont
  {de~Vito}}, \bibinfo {author} {\bibfnamefont {C.}~\bibnamefont {Fornetto}},
  \bibinfo {author} {\bibfnamefont {P.}~\bibnamefont {Ricci}}, \bibinfo
  {author} {\bibfnamefont {C.}~\bibnamefont {M{\"u}llenbroich}}, \bibinfo
  {author} {\bibfnamefont {G.}~\bibnamefont {Sancataldo}}, \bibinfo {author}
  {\bibfnamefont {L.}~\bibnamefont {Turrini}}, \bibinfo {author} {\bibfnamefont
  {G.}~\bibnamefont {Mazzamuto}}, \bibinfo {author} {\bibfnamefont
  {N.}~\bibnamefont {Tiso}}, \bibinfo {author} {\bibfnamefont {L.}~\bibnamefont
  {Sacconi}}, \bibinfo {author} {\bibfnamefont {D.}~\bibnamefont {Fanelli}},
  \emph {et~al.},\ }in\ \href@noop {} {\emph {\bibinfo {booktitle} {Neural
  Imaging and Sensing 2020}}},\ Vol.\ \bibinfo {volume} {11226}\ (\bibinfo
  {organization} {International Society for Optics and Photonics},\ \bibinfo
  {year} {2020})\ p.\ \bibinfo {pages} {1122604}\BibitemShut {NoStop}%
\bibitem [{\citenamefont {de~Vito}\ \emph
  {et~al.}(2020{\natexlab{b}})\citenamefont {de~Vito}, \citenamefont {Turrini},
  \citenamefont {Fornetto}, \citenamefont {Ricci}, \citenamefont
  {M{\"u}llenbroich}, \citenamefont {Sancataldo}, \citenamefont {Trabalzini},
  \citenamefont {Mazzamuto}, \citenamefont {Tiso}, \citenamefont {Sacconi}
  \emph {et~al.}}]{de2020twob}%
  \BibitemOpen
  \bibfield  {author} {\bibinfo {author} {\bibfnamefont {G.}~\bibnamefont
  {de~Vito}}, \bibinfo {author} {\bibfnamefont {L.}~\bibnamefont {Turrini}},
  \bibinfo {author} {\bibfnamefont {C.}~\bibnamefont {Fornetto}}, \bibinfo
  {author} {\bibfnamefont {P.}~\bibnamefont {Ricci}}, \bibinfo {author}
  {\bibfnamefont {C.}~\bibnamefont {M{\"u}llenbroich}}, \bibinfo {author}
  {\bibfnamefont {G.}~\bibnamefont {Sancataldo}}, \bibinfo {author}
  {\bibfnamefont {E.}~\bibnamefont {Trabalzini}}, \bibinfo {author}
  {\bibfnamefont {G.}~\bibnamefont {Mazzamuto}}, \bibinfo {author}
  {\bibfnamefont {N.}~\bibnamefont {Tiso}}, \bibinfo {author} {\bibfnamefont
  {L.}~\bibnamefont {Sacconi}},  \emph {et~al.},\ }in\ \href@noop {} {\emph
  {\bibinfo {booktitle} {Neurophotonics}}},\ Vol.\ \bibinfo {volume} {11360}\
  (\bibinfo {organization} {International Society for Optics and Photonics},\
  \bibinfo {year} {2020})\ p.\ \bibinfo {pages} {1136004}\BibitemShut {NoStop}%
\bibitem [{\citenamefont {Randlett}\ \emph {et~al.}(2015)\citenamefont
  {Randlett}, \citenamefont {Wee}, \citenamefont {Naumann}, \citenamefont
  {Nnaemeka}, \citenamefont {Schoppik}, \citenamefont {Fitzgerald},
  \citenamefont {Portugues}, \citenamefont {Lacoste}, \citenamefont {Riegler},
  \citenamefont {Engert} \emph {et~al.}}]{randlett2015whole}%
  \BibitemOpen
  \bibfield  {author} {\bibinfo {author} {\bibfnamefont {O.}~\bibnamefont
  {Randlett}}, \bibinfo {author} {\bibfnamefont {C.~L.}\ \bibnamefont {Wee}},
  \bibinfo {author} {\bibfnamefont {E.~A.}\ \bibnamefont {Naumann}}, \bibinfo
  {author} {\bibfnamefont {O.}~\bibnamefont {Nnaemeka}}, \bibinfo {author}
  {\bibfnamefont {D.}~\bibnamefont {Schoppik}}, \bibinfo {author}
  {\bibfnamefont {J.~E.}\ \bibnamefont {Fitzgerald}}, \bibinfo {author}
  {\bibfnamefont {R.}~\bibnamefont {Portugues}}, \bibinfo {author}
  {\bibfnamefont {A.~M.}\ \bibnamefont {Lacoste}}, \bibinfo {author}
  {\bibfnamefont {C.}~\bibnamefont {Riegler}}, \bibinfo {author} {\bibfnamefont
  {F.}~\bibnamefont {Engert}},  \emph {et~al.},\ }\href@noop {} {\bibfield
  {journal} {\bibinfo  {journal} {Nature Methods}\ }\textbf {\bibinfo {volume}
  {12}},\ \bibinfo {pages} {1039} (\bibinfo {year} {2015})}\BibitemShut
  {NoStop}%
\bibitem [{\citenamefont {Kunst}\ \emph {et~al.}(2019)\citenamefont {Kunst},
  \citenamefont {Laurell}, \citenamefont {Mokayes}, \citenamefont {Kramer},
  \citenamefont {Kubo}, \citenamefont {Fernandes}, \citenamefont {F{\"o}rster},
  \citenamefont {Dal~Maschio},\ and\ \citenamefont
  {Baier}}]{kunst2019cellular}%
  \BibitemOpen
  \bibfield  {author} {\bibinfo {author} {\bibfnamefont {M.}~\bibnamefont
  {Kunst}}, \bibinfo {author} {\bibfnamefont {E.}~\bibnamefont {Laurell}},
  \bibinfo {author} {\bibfnamefont {N.}~\bibnamefont {Mokayes}}, \bibinfo
  {author} {\bibfnamefont {A.}~\bibnamefont {Kramer}}, \bibinfo {author}
  {\bibfnamefont {F.}~\bibnamefont {Kubo}}, \bibinfo {author} {\bibfnamefont
  {A.~M.}\ \bibnamefont {Fernandes}}, \bibinfo {author} {\bibfnamefont
  {D.}~\bibnamefont {F{\"o}rster}}, \bibinfo {author} {\bibfnamefont
  {M.}~\bibnamefont {Dal~Maschio}}, \ and\ \bibinfo {author} {\bibfnamefont
  {H.}~\bibnamefont {Baier}},\ }\href@noop {} {\bibfield  {journal} {\bibinfo
  {journal} {Neuron}\ }\textbf {\bibinfo {volume} {103}},\ \bibinfo {pages}
  {21} (\bibinfo {year} {2019})}\BibitemShut {NoStop}%
\bibitem [{\citenamefont {Tsodyks}\ \emph {et~al.}(1998)\citenamefont
  {Tsodyks}, \citenamefont {Pawelzik},\ and\ \citenamefont {Markram}}]{TUM1}%
  \BibitemOpen
  \bibfield  {author} {\bibinfo {author} {\bibfnamefont {M.}~\bibnamefont
  {Tsodyks}}, \bibinfo {author} {\bibfnamefont {K.}~\bibnamefont {Pawelzik}}, \
  and\ \bibinfo {author} {\bibfnamefont {H.}~\bibnamefont {Markram}},\
  }\href@noop {} {\bibfield  {journal} {\bibinfo  {journal} {Neural
  Computation}\ }\textbf {\bibinfo {volume} {10}},\ \bibinfo {pages} {821}
  (\bibinfo {year} {1998})}\BibitemShut {NoStop}%
\bibitem [{\citenamefont {Tsodyks}\ and\ \citenamefont {Markram}(1997)}]{TUM2}%
  \BibitemOpen
  \bibfield  {author} {\bibinfo {author} {\bibfnamefont {M.~V.}\ \bibnamefont
  {Tsodyks}}\ and\ \bibinfo {author} {\bibfnamefont {H.}~\bibnamefont
  {Markram}},\ }\href@noop {} {\bibfield  {journal} {\bibinfo  {journal}
  {Proceedings of the National Academy of Sciences}\ }\textbf {\bibinfo
  {volume} {94}},\ \bibinfo {pages} {719} (\bibinfo {year} {1997})}\BibitemShut
  {NoStop}%
\bibitem [{\citenamefont {Tsodyks}\ \emph {et~al.}(2000)\citenamefont
  {Tsodyks}, \citenamefont {Uziel},\ and\ \citenamefont {Markram}}]{TUM}%
  \BibitemOpen
  \bibfield  {author} {\bibinfo {author} {\bibfnamefont {M.}~\bibnamefont
  {Tsodyks}}, \bibinfo {author} {\bibfnamefont {A.}~\bibnamefont {Uziel}}, \
  and\ \bibinfo {author} {\bibfnamefont {H.}~\bibnamefont {Markram}},\
  }\href@noop {} {\bibfield  {journal} {\bibinfo  {journal} {The Journal of
  Neuroscience}\ }\textbf {\bibinfo {volume} {20}},\ \bibinfo {pages} {825}
  (\bibinfo {year} {2000})}\BibitemShut {NoStop}%
\bibitem [{\citenamefont {Baker}\ \emph {et~al.}(1971)\citenamefont {Baker},
  \citenamefont {Hodgkin},\ and\ \citenamefont
  {Ridgway}}]{baker1971depolarization}%
  \BibitemOpen
  \bibfield  {author} {\bibinfo {author} {\bibfnamefont {P.}~\bibnamefont
  {Baker}}, \bibinfo {author} {\bibfnamefont {A.}~\bibnamefont {Hodgkin}}, \
  and\ \bibinfo {author} {\bibfnamefont {E.}~\bibnamefont {Ridgway}},\
  }\href@noop {} {\bibfield  {journal} {\bibinfo  {journal} {The Journal of
  Physiology}\ }\textbf {\bibinfo {volume} {218}},\ \bibinfo {pages} {709}
  (\bibinfo {year} {1971})}\BibitemShut {NoStop}%
\bibitem [{\citenamefont {Grienberger}\ and\ \citenamefont
  {Konnerth}(2012)}]{grienberger2012imaging}%
  \BibitemOpen
  \bibfield  {author} {\bibinfo {author} {\bibfnamefont {C.}~\bibnamefont
  {Grienberger}}\ and\ \bibinfo {author} {\bibfnamefont {A.}~\bibnamefont
  {Konnerth}},\ }\href@noop {} {\bibfield  {journal} {\bibinfo  {journal}
  {Neuron}\ }\textbf {\bibinfo {volume} {73}},\ \bibinfo {pages} {862}
  (\bibinfo {year} {2012})}\BibitemShut {NoStop}%
\bibitem [{\citenamefont {Naumann}\ \emph {et~al.}(2010)\citenamefont
  {Naumann}, \citenamefont {Kampff}, \citenamefont {Prober}, \citenamefont
  {Schier},\ and\ \citenamefont {Engert}}]{naumann2010monitoring}%
  \BibitemOpen
  \bibfield  {author} {\bibinfo {author} {\bibfnamefont {E.~A.}\ \bibnamefont
  {Naumann}}, \bibinfo {author} {\bibfnamefont {A.~R.}\ \bibnamefont {Kampff}},
  \bibinfo {author} {\bibfnamefont {D.~A.}\ \bibnamefont {Prober}}, \bibinfo
  {author} {\bibfnamefont {A.~F.}\ \bibnamefont {Schier}}, \ and\ \bibinfo
  {author} {\bibfnamefont {F.}~\bibnamefont {Engert}},\ }\href@noop {}
  {\bibfield  {journal} {\bibinfo  {journal} {Nature Neuroscience}\ }\textbf
  {\bibinfo {volume} {13}},\ \bibinfo {pages} {513} (\bibinfo {year}
  {2010})}\BibitemShut {NoStop}%
\bibitem [{\citenamefont {Chen}\ \emph {et~al.}(2013)\citenamefont {Chen},
  \citenamefont {Wardill}, \citenamefont {Sun}, \citenamefont {Pulver},
  \citenamefont {Renninger}, \citenamefont {Baohan}, \citenamefont {Schreiter},
  \citenamefont {Kerr}, \citenamefont {Orger}, \citenamefont {Jayaraman},
  \citenamefont {Looger}, \citenamefont {Svoboda},\ and\ \citenamefont
  {Kim}}]{Chen2013}%
  \BibitemOpen
  \bibfield  {author} {\bibinfo {author} {\bibfnamefont {T.-W.}\ \bibnamefont
  {Chen}}, \bibinfo {author} {\bibfnamefont {T.~J.}\ \bibnamefont {Wardill}},
  \bibinfo {author} {\bibfnamefont {Y.}~\bibnamefont {Sun}}, \bibinfo {author}
  {\bibfnamefont {S.~R.}\ \bibnamefont {Pulver}}, \bibinfo {author}
  {\bibfnamefont {S.~L.}\ \bibnamefont {Renninger}}, \bibinfo {author}
  {\bibfnamefont {A.}~\bibnamefont {Baohan}}, \bibinfo {author} {\bibfnamefont
  {E.~R.}\ \bibnamefont {Schreiter}}, \bibinfo {author} {\bibfnamefont {R.~A.}\
  \bibnamefont {Kerr}}, \bibinfo {author} {\bibfnamefont {M.~B.}\ \bibnamefont
  {Orger}}, \bibinfo {author} {\bibfnamefont {V.}~\bibnamefont {Jayaraman}},
  \bibinfo {author} {\bibfnamefont {L.~L.}\ \bibnamefont {Looger}}, \bibinfo
  {author} {\bibfnamefont {K.}~\bibnamefont {Svoboda}}, \ and\ \bibinfo
  {author} {\bibfnamefont {D.~S.}\ \bibnamefont {Kim}},\ }\href {\doibase
  10.1038/nature12354} {\bibfield  {journal} {\bibinfo  {journal} {Nature}\
  }\textbf {\bibinfo {volume} {499}},\ \bibinfo {pages} {295} (\bibinfo {year}
  {2013})}\BibitemShut {NoStop}%
\bibitem [{\citenamefont {Volman}\ \emph {et~al.}(2004)\citenamefont {Volman},
  \citenamefont {Baruchi}, \citenamefont {Persi},\ and\ \citenamefont
  {Ben-Jacob}}]{volman}%
  \BibitemOpen
  \bibfield  {author} {\bibinfo {author} {\bibfnamefont {V.}~\bibnamefont
  {Volman}}, \bibinfo {author} {\bibfnamefont {I.}~\bibnamefont {Baruchi}},
  \bibinfo {author} {\bibfnamefont {E.}~\bibnamefont {Persi}}, \ and\ \bibinfo
  {author} {\bibfnamefont {E.}~\bibnamefont {Ben-Jacob}},\ }\href@noop {}
  {\bibfield  {journal} {\bibinfo  {journal} {Physica A: Statistical Mechanics
  and its Applications}\ }\textbf {\bibinfo {volume} {335}},\ \bibinfo {pages}
  {249} (\bibinfo {year} {2004})}\BibitemShut {NoStop}%
\bibitem [{\citenamefont {Northcutt}(2008)}]{northcutt2008forebrain}%
  \BibitemOpen
  \bibfield  {author} {\bibinfo {author} {\bibfnamefont {R.~G.}\ \bibnamefont
  {Northcutt}},\ }\href@noop {} {\bibfield  {journal} {\bibinfo  {journal}
  {Brain Research Bulletin}\ }\textbf {\bibinfo {volume} {75}},\ \bibinfo
  {pages} {191} (\bibinfo {year} {2008})}\BibitemShut {NoStop}%
\bibitem [{\citenamefont {Mueller}(2012)}]{mueller2012thalamus}%
  \BibitemOpen
  \bibfield  {author} {\bibinfo {author} {\bibfnamefont {T.}~\bibnamefont
  {Mueller}},\ }\href@noop {} {\bibfield  {journal} {\bibinfo  {journal}
  {Frontiers in Neuroscience}\ }\textbf {\bibinfo {volume} {6}},\ \bibinfo
  {pages} {64} (\bibinfo {year} {2012})}\BibitemShut {NoStop}%
\bibitem [{\citenamefont {Yokogawa}\ \emph {et~al.}(2012)\citenamefont
  {Yokogawa}, \citenamefont {Hannan},\ and\ \citenamefont
  {Burgess}}]{yokogawa2012dorsal}%
  \BibitemOpen
  \bibfield  {author} {\bibinfo {author} {\bibfnamefont {T.}~\bibnamefont
  {Yokogawa}}, \bibinfo {author} {\bibfnamefont {M.~C.}\ \bibnamefont
  {Hannan}}, \ and\ \bibinfo {author} {\bibfnamefont {H.~A.}\ \bibnamefont
  {Burgess}},\ }\href@noop {} {\bibfield  {journal} {\bibinfo  {journal} {The
  Journal of Neuroscience}\ }\textbf {\bibinfo {volume} {32}},\ \bibinfo
  {pages} {15205} (\bibinfo {year} {2012})}\BibitemShut {NoStop}%
\bibitem [{\citenamefont {Garcia-Campmany}\ \emph {et~al.}(2010)\citenamefont
  {Garcia-Campmany}, \citenamefont {Stam},\ and\ \citenamefont
  {Goulding}}]{garcia2010circuits}%
  \BibitemOpen
  \bibfield  {author} {\bibinfo {author} {\bibfnamefont {L.}~\bibnamefont
  {Garcia-Campmany}}, \bibinfo {author} {\bibfnamefont {F.~J.}\ \bibnamefont
  {Stam}}, \ and\ \bibinfo {author} {\bibfnamefont {M.}~\bibnamefont
  {Goulding}},\ }\href@noop {} {\bibfield  {journal} {\bibinfo  {journal}
  {Current Opinion in Neurobiology}\ }\textbf {\bibinfo {volume} {20}},\
  \bibinfo {pages} {116} (\bibinfo {year} {2010})}\BibitemShut {NoStop}%
\bibitem [{\citenamefont {Kinkhabwala}\ \emph {et~al.}(2011)\citenamefont
  {Kinkhabwala}, \citenamefont {Riley}, \citenamefont {Koyama}, \citenamefont
  {Monen}, \citenamefont {Satou}, \citenamefont {Kimura}, \citenamefont
  {Higashijima},\ and\ \citenamefont {Fetcho}}]{kinkhabwala2011structural}%
  \BibitemOpen
  \bibfield  {author} {\bibinfo {author} {\bibfnamefont {A.}~\bibnamefont
  {Kinkhabwala}}, \bibinfo {author} {\bibfnamefont {M.}~\bibnamefont {Riley}},
  \bibinfo {author} {\bibfnamefont {M.}~\bibnamefont {Koyama}}, \bibinfo
  {author} {\bibfnamefont {J.}~\bibnamefont {Monen}}, \bibinfo {author}
  {\bibfnamefont {C.}~\bibnamefont {Satou}}, \bibinfo {author} {\bibfnamefont
  {Y.}~\bibnamefont {Kimura}}, \bibinfo {author} {\bibfnamefont {S.-i.}\
  \bibnamefont {Higashijima}}, \ and\ \bibinfo {author} {\bibfnamefont
  {J.}~\bibnamefont {Fetcho}},\ }\href@noop {} {\bibfield  {journal} {\bibinfo
  {journal} {Proceedings of the National Academy of Sciences}\ }\textbf
  {\bibinfo {volume} {108}},\ \bibinfo {pages} {1164} (\bibinfo {year}
  {2011})}\BibitemShut {NoStop}%
\bibitem [{\citenamefont {Heap}\ \emph {et~al.}(2013)\citenamefont {Heap},
  \citenamefont {Goh}, \citenamefont {Kassahn},\ and\ \citenamefont
  {Scott}}]{heap2013cerebellar}%
  \BibitemOpen
  \bibfield  {author} {\bibinfo {author} {\bibfnamefont {L.}~\bibnamefont
  {Heap}}, \bibinfo {author} {\bibfnamefont {C.-C.}\ \bibnamefont {Goh}},
  \bibinfo {author} {\bibfnamefont {K.~S.}\ \bibnamefont {Kassahn}}, \ and\
  \bibinfo {author} {\bibfnamefont {E.~K.}\ \bibnamefont {Scott}},\ }\href@noop
  {} {\bibfield  {journal} {\bibinfo  {journal} {Frontiers in Neural Circuits}\
  }\textbf {\bibinfo {volume} {7}},\ \bibinfo {pages} {53} (\bibinfo {year}
  {2013})}\BibitemShut {NoStop}%
\bibitem [{\citenamefont {Kaslin}\ and\ \citenamefont
  {Brand}(2016)}]{kaslin2016zebrafish}%
  \BibitemOpen
  \bibfield  {author} {\bibinfo {author} {\bibfnamefont {J.}~\bibnamefont
  {Kaslin}}\ and\ \bibinfo {author} {\bibfnamefont {M.}~\bibnamefont {Brand}},\
  }in\ \href@noop {} {\emph {\bibinfo {booktitle} {Essentials of Cerebellum and
  Cerebellar Disorders}}}\ (\bibinfo  {publisher} {Springer},\ \bibinfo {year}
  {2016})\ pp.\ \bibinfo {pages} {411--421}\BibitemShut {NoStop}%
\bibitem [{\citenamefont {Sajovic}\ and\ \citenamefont
  {Levinthal}(1982)}]{sajovic1982visual}%
  \BibitemOpen
  \bibfield  {author} {\bibinfo {author} {\bibfnamefont {P.}~\bibnamefont
  {Sajovic}}\ and\ \bibinfo {author} {\bibfnamefont {C.}~\bibnamefont
  {Levinthal}},\ }\href@noop {} {\bibfield  {journal} {\bibinfo  {journal}
  {Neuroscience}\ }\textbf {\bibinfo {volume} {7}},\ \bibinfo {pages} {2407}
  (\bibinfo {year} {1982})}\BibitemShut {NoStop}%
\bibitem [{\citenamefont {Gebhardt}\ \emph {et~al.}(2019)\citenamefont
  {Gebhardt}, \citenamefont {Auer}, \citenamefont {Henriques}, \citenamefont
  {Rajan}, \citenamefont {Duroure}, \citenamefont {Bianco},\ and\ \citenamefont
  {Del~Bene}}]{gebhardt2019interhemispheric}%
  \BibitemOpen
  \bibfield  {author} {\bibinfo {author} {\bibfnamefont {C.}~\bibnamefont
  {Gebhardt}}, \bibinfo {author} {\bibfnamefont {T.~O.}\ \bibnamefont {Auer}},
  \bibinfo {author} {\bibfnamefont {P.~M.}\ \bibnamefont {Henriques}}, \bibinfo
  {author} {\bibfnamefont {G.}~\bibnamefont {Rajan}}, \bibinfo {author}
  {\bibfnamefont {K.}~\bibnamefont {Duroure}}, \bibinfo {author} {\bibfnamefont
  {I.~H.}\ \bibnamefont {Bianco}}, \ and\ \bibinfo {author} {\bibfnamefont
  {F.}~\bibnamefont {Del~Bene}},\ }\href@noop {} {\bibfield  {journal}
  {\bibinfo  {journal} {Nature Communications}\ }\textbf {\bibinfo {volume}
  {10}},\ \bibinfo {pages} {1} (\bibinfo {year} {2019})}\BibitemShut {NoStop}%
\bibitem [{\citenamefont {de~Vito}\ \emph
  {et~al.}(2020{\natexlab{c}})\citenamefont {de~Vito}, \citenamefont {Ricci},
  \citenamefont {Turrini}, \citenamefont {Tiso}, \citenamefont {Vanzi},
  \citenamefont {Silvestri},\ and\ \citenamefont {Pavone}}]{de2020effects}%
  \BibitemOpen
  \bibfield  {author} {\bibinfo {author} {\bibfnamefont {G.}~\bibnamefont
  {de~Vito}}, \bibinfo {author} {\bibfnamefont {P.}~\bibnamefont {Ricci}},
  \bibinfo {author} {\bibfnamefont {L.}~\bibnamefont {Turrini}}, \bibinfo
  {author} {\bibfnamefont {N.}~\bibnamefont {Tiso}}, \bibinfo {author}
  {\bibfnamefont {F.}~\bibnamefont {Vanzi}}, \bibinfo {author} {\bibfnamefont
  {L.}~\bibnamefont {Silvestri}}, \ and\ \bibinfo {author} {\bibfnamefont
  {F.~S.}\ \bibnamefont {Pavone}},\ }\href@noop {} {\bibfield  {journal}
  {\bibinfo  {journal} {arXiv preprint arXiv:2003.09496}\ } (\bibinfo {year}
  {2020}{\natexlab{c}})}\BibitemShut {NoStop}%
\bibitem [{\citenamefont {Vladimirov}\ \emph {et~al.}(2014)\citenamefont
  {Vladimirov}, \citenamefont {Mu}, \citenamefont {Kawashima}, \citenamefont
  {Bennett}, \citenamefont {Yang}, \citenamefont {Looger}, \citenamefont
  {Keller}, \citenamefont {Freeman},\ and\ \citenamefont
  {Ahrens}}]{vladimirov2014light}%
  \BibitemOpen
  \bibfield  {author} {\bibinfo {author} {\bibfnamefont {N.}~\bibnamefont
  {Vladimirov}}, \bibinfo {author} {\bibfnamefont {Y.}~\bibnamefont {Mu}},
  \bibinfo {author} {\bibfnamefont {T.}~\bibnamefont {Kawashima}}, \bibinfo
  {author} {\bibfnamefont {D.~V.}\ \bibnamefont {Bennett}}, \bibinfo {author}
  {\bibfnamefont {C.-T.}\ \bibnamefont {Yang}}, \bibinfo {author}
  {\bibfnamefont {L.~L.}\ \bibnamefont {Looger}}, \bibinfo {author}
  {\bibfnamefont {P.~J.}\ \bibnamefont {Keller}}, \bibinfo {author}
  {\bibfnamefont {J.}~\bibnamefont {Freeman}}, \ and\ \bibinfo {author}
  {\bibfnamefont {M.~B.}\ \bibnamefont {Ahrens}},\ }\href@noop {} {\bibfield
  {journal} {\bibinfo  {journal} {Nature Methods}\ }\textbf {\bibinfo {volume}
  {11}},\ \bibinfo {pages} {883} (\bibinfo {year} {2014})}\BibitemShut
  {NoStop}%
\bibitem [{\citenamefont {M{\"u}llenbroich}\ \emph {et~al.}(2018)\citenamefont
  {M{\"u}llenbroich}, \citenamefont {Turrini}, \citenamefont {Silvestri},
  \citenamefont {Alterini}, \citenamefont {Gheisari}, \citenamefont {Tiso},
  \citenamefont {Vanzi}, \citenamefont {Sacconi},\ and\ \citenamefont
  {Pavone}}]{mullenbroich2018bessel}%
  \BibitemOpen
  \bibfield  {author} {\bibinfo {author} {\bibfnamefont {M.~C.}\ \bibnamefont
  {M{\"u}llenbroich}}, \bibinfo {author} {\bibfnamefont {L.}~\bibnamefont
  {Turrini}}, \bibinfo {author} {\bibfnamefont {L.}~\bibnamefont {Silvestri}},
  \bibinfo {author} {\bibfnamefont {T.}~\bibnamefont {Alterini}}, \bibinfo
  {author} {\bibfnamefont {A.}~\bibnamefont {Gheisari}}, \bibinfo {author}
  {\bibfnamefont {N.}~\bibnamefont {Tiso}}, \bibinfo {author} {\bibfnamefont
  {F.}~\bibnamefont {Vanzi}}, \bibinfo {author} {\bibfnamefont
  {L.}~\bibnamefont {Sacconi}}, \ and\ \bibinfo {author} {\bibfnamefont
  {F.~S.}\ \bibnamefont {Pavone}},\ }\href@noop {} {\bibfield  {journal}
  {\bibinfo  {journal} {Frontiers in Cellular Neuroscience}\ }\textbf {\bibinfo
  {volume} {12}},\ \bibinfo {pages} {315} (\bibinfo {year} {2018})}\BibitemShut
  {NoStop}%
\bibitem [{\citenamefont {Turrini}\ \emph {et~al.}(2017)\citenamefont
  {Turrini}, \citenamefont {Fornetto}, \citenamefont {Marchetto}, \citenamefont
  {M{\"u}llenbroich}, \citenamefont {Tiso}, \citenamefont {Vettori},
  \citenamefont {Resta}, \citenamefont {Masi}, \citenamefont {Mannaioni},
  \citenamefont {Pavone} \emph {et~al.}}]{turrini2017optical}%
  \BibitemOpen
  \bibfield  {author} {\bibinfo {author} {\bibfnamefont {L.}~\bibnamefont
  {Turrini}}, \bibinfo {author} {\bibfnamefont {C.}~\bibnamefont {Fornetto}},
  \bibinfo {author} {\bibfnamefont {G.}~\bibnamefont {Marchetto}}, \bibinfo
  {author} {\bibfnamefont {M.}~\bibnamefont {M{\"u}llenbroich}}, \bibinfo
  {author} {\bibfnamefont {N.}~\bibnamefont {Tiso}}, \bibinfo {author}
  {\bibfnamefont {A.}~\bibnamefont {Vettori}}, \bibinfo {author} {\bibfnamefont
  {F.}~\bibnamefont {Resta}}, \bibinfo {author} {\bibfnamefont
  {A.}~\bibnamefont {Masi}}, \bibinfo {author} {\bibfnamefont {G.}~\bibnamefont
  {Mannaioni}}, \bibinfo {author} {\bibfnamefont {F.}~\bibnamefont {Pavone}},
  \emph {et~al.},\ }\href@noop {} {\bibfield  {journal} {\bibinfo  {journal}
  {Scientific Reports}\ }\textbf {\bibinfo {volume} {7}},\ \bibinfo {pages} {1}
  (\bibinfo {year} {2017})}\BibitemShut {NoStop}%
\bibitem [{\citenamefont {Westerfield}(2000)}]{westerfield2000zebrafish}%
  \BibitemOpen
  \bibfield  {author} {\bibinfo {author} {\bibfnamefont {M.}~\bibnamefont
  {Westerfield}},\ }\href@noop {} {\bibfield  {journal} {\bibinfo  {journal}
  {http://zfin. org/zf\_info/zfbook/zfbk. html}\ } (\bibinfo {year}
  {2000})}\BibitemShut {NoStop}%
\end{thebibliography}%

\end{document}